\documentclass{article}

\usepackage{arxiv}

\usepackage[utf8]{inputenc} 
\usepackage[T1]{fontenc}    
\usepackage{hyperref}       
\usepackage{url}            
\usepackage{booktabs}       
\usepackage{amsfonts}       
\usepackage{nicefrac}       
\usepackage{microtype}      
\usepackage{lipsum}
\usepackage{xcolor}

\usepackage[title]{appendix}

\usepackage{caption}
\usepackage{subcaption}

\usepackage{enumitem}
\usepackage{float}
\usepackage{dblfloatfix}
\usepackage{titlesec}

\usepackage{array}
\usepackage{mdwmath}
\usepackage{mdwtab}

\usepackage{amsmath,amsfonts,amsthm} 
\usepackage{mathtools}

\usepackage{graphicx}
\usepackage{listings}
\usepackage{algorithm2e}

\usepackage{changepage}
\usepackage[utf8]{inputenc}

\newcommand\nnfootnote[1]{%
  \begin{NoHyper}
  \renewcommand\thefootnote{}\footnote{#1}%
  \addtocounter{footnote}{-1}%
  \end{NoHyper}
}

\usepackage[user,titleref]{zref}

\usepackage{ulem}

\usepackage[auth-sc]{authblk}

\usepackage[misc, geometry]{ifsym}

\title{Opinion Manipulation on Farsi Twitter}

\author[1*,2,3]{Amirhossein Farzam}
\author[4,5]{Parham Moradi}
\author[4,5, 6*]{Saeedeh Mohammadi}
\author[7]{Zahra Padar}
\author[8]{Alexandra A. Siegel}
\affil[1]{Department of Political Science, Duke University, Durham, NC 27708}
\affil[2]{Department of Engineering Sciences and Applied Mathematics, Northwestern University, Evanston, IL 60208}
\affil[3]{Max Planck Institute for Mathematics in the Sciences, 04103 Leipzig, Germany}
\affil[4]{Center for Complex Networks and Social Data Science, Tehran, Iran}
\affil[5]{Department of Physics, Shahid Beheshti University, Tehran, Iran}
\affil[6]{School of Mathematics and Statistics, University College Dublin, Belfield, Dublin 4, Ireland}
\affil[7]{Department of Computer Science, University of Freiburg, Freiburg, Germany}
\affil[8]{Department of Political Science, University of Colorado, Boulder, CO 80309}

\begin{document}

\maketitle

\begin{abstract}
For Iranians and the Iranian diaspora, the Farsi Twittersphere provides an important alternative to state media and an outlet for political discourse. But this understudied online space has become an opinion manipulation battleground, with diverse actors using inauthentic accounts to advance their goals and shape online narratives. Examining trending discussions crossing social cleavages in Iran, we explore how the dynamics of opinion manipulation differ across diverse issue areas. Our analysis suggests that opinion manipulation by inauthentic accounts is more prevalent in divisive political discussions than non-divisive or apolitical discussions. 
We show how Twitter’s network structures help to reinforce the content propagated by clusters of inauthentic accounts in divisive political discussions. 
Analyzing both the content and structure of online discussions in the Iranian Twittersphere, this work contributes to a growing body of literature exploring the dynamics of online opinion manipulation, while improving our understanding of how information is controlled in the digital age. 
\end{abstract}


\nnfootnote{Corresponding authors: a.farzam@duke.edu, Alexandra.Siegel@colorado.edu}
\nnfootnote{* Current affiliation}

\linespread{1.25}\selectfont


\section*{Introduction}\zlabel{section_introduction}

The Farsi Twittersphere was initially hailed as an early example of ``liberation technology'' for its role in the 2009 anti-regime protests in Iran \cite{diamond2010liberation}. 
In the absence of independent media, Farsi Twitter emerged as a public space where uncensored opinions proliferated \cite{Ketabchi_postelectionTwitter}. 
But over the past decade, the Farsi Twittersphere has become an opinion manipulation battleground. Diverse actors regularly employ bots, sock-puppets, and other inauthentic accounts to advance their goals and shape online narratives \cite{Thieltges_socialBotsIran}. 
Examining over one million tweets collected between June 2019 and January 2020, we explore the behavior of inauthentic accounts across trending online discussions spanning diverse social and political issues in the Farsi Twittersphere.  

The influence of coordinated inauthentic activity in online political debates is a topic of growing interest among academics and policymakers alike. 
An emerging body of research has highlighted the considerable presence of inauthentic accounts working to shape debates around elections, political protests, military operations, and other political events in diverse contexts. \cite{Ananyev_FantasticBeasts, Bessi_SocialBotsUSElection, Caldarelli_botSquads, Chu_WhoIsTweeting, Forelle_botsVenezuela, Gorwa_typologyGuideToSocialBots, Howard_botsBrexit, Shao_spreadOfFakeNews, Schaefer_JapansGeneralElection, Tucker_reviewSocialMediaBots, Woolley_socialBotInterference}). 
Such accounts include bots, cyborgs, sock-puppets, and trolls (e.g. \cite{Ananyev_FantasticBeasts, Chu_WhoIsTweeting, Gorwa_typologyGuideToSocialBots, Tucker_reviewSocialMediaBots}). 
The majority of this work explores the use of inauthentic activity by foreign actors to influence politics abroad, with Russian interference in US elections receiving the most scholarly attention. 
The use of inauthentic accounts by domestic actors to influence domestic politics is particularly understudied (though see \cite{barrie2021kingdom} for an exception). 

We assess the relative prevalence of inauthentic activity in eleven trending Twitter discussion topics crossing the social cleavages of the Iranian society. 
Specifically, we introduce a typology of authentic and inauthentic Twitter users and examine how different types of users participate in trending discussions over time. 
To classify users according to our typology, we use Botometer \cite{Davis_botOrNot, Varol_Botometer, Yang_Botometer}, a supervised machine-learning algorithm, to measure the Complete Automation Probability (CAP) for the users in our Twitter data. 
While Botometer does not accurately classify individual accounts, extensive human coding and validation using Farsi Twitter accounts suggests that average CAP scores can be used to assess the average relative prevalence of inauthentic activity across communities or groups of Twitter users.

After developing these relative measures of inauthenticity, we analyze friendship and retweet networks to understand how clusters of inauthentic accounts interact with diverse communities of Iranian Twitter users across the political spectrum. 
Exploring the network structures of Twitter users across these trending discussions, we find that inauthentic activity is concentrated in divisive political discussions, where retweet and friendship networks are ideologically polarized. By contrast, apolitical discussions have very little inauthentic activity, and these networks are characterized by cross cutting interactions across the political spectrum. 

Finally, we compare the content produced by relatively inauthentic groups of accounts to more genuine clusters of accounts across different trending topics. 
We find that clusters of inauthentic users in divisive political discussions use highly polarizing keywords, targeting particular ideological audiences, relative to clusters of authentic users. 
Characterizing both the content and structure of online discussions in the Iranian Twittersphere across diverse issue areas, this work contributes to a growing body of literature exploring the dynamics of online opinion manipulation.

\section*{Results}\zlabel{section_Results}

\subsection*{Prevalence of Inauthentic Accounts}
\zlabel{section_results_subsection_prevalence}

\begin{figure}[!htb]
	\centering
	\captionsetup{width=.9\linewidth, format=hang}
	\includegraphics[width=.5\linewidth]{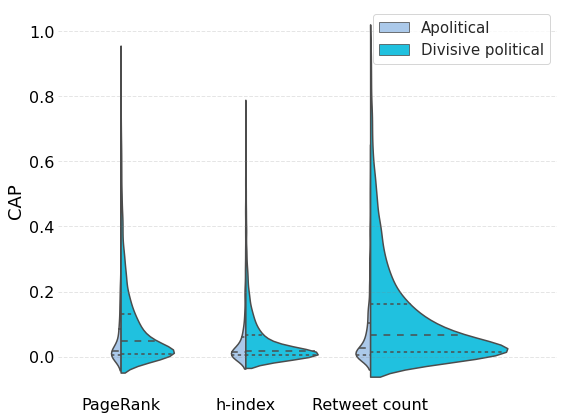}
	\caption{ \footnotesize 
	The distributions of CAP scores among the most influential accounts, by PageRank (left), h-index (middle), and retweet count (right), in the retweet network of the Apolitical and Divisive Political discussions.
	The violin plots, and the quartile lines within each violin plot, show the distribution of CAP scores for users among the top decile of accounts with respect to the corresponding measure, in each topic within each category of discussion.
	The two-sided Kolmogorov-Smirnov test reveals statistically significant difference between each pair of distributions at significance level $p < 0.001$ for PageRank and retweet count, but this difference is not statistically significant for h-index.
	The left half of the violin plot (Apolitical) contains 790 CAP scores and the right half (Divisive Political) contains 7385 CAP scores.}
	\captionsetup{labelformat=empty}
	\label{fig:fig_ADvsDPD_cap_dist}
\end{figure}

To assess the role of inauthentic accounts in trending topics in the Farsi Twittersphere, we collected data in real time surrounding 11 trending topics---four apolitical discussions, four non-divisive political discussions, and three divisive political discussions between June 2019 and January 2020.
Developing a typology of Twitter users with respect to the level of automation and inauthenticity of the account, we examine the average relative prevalence of different types of accounts in each trending discussion.
We use the CAP scores from Botometer as a relative measure of the level of inauthenticity of the accounts.
The details of how CAP scores are utilized for this purpose are explained in the \ztitleref{section_Methods_subsection_typology} subsection.
As Fig. \ref{fig:fig_ADvsDPD_cap_dist} displays, we see that {most influential accounts---the top decile of most retweeted accounts---}in divisive political discussions, on average {tend to} have higher levels of inauthentic activity (higher CAP scores) than non divisive political discussions and apolitical discussions. 
Performing a two-sided two-sample Kolmogorov-Smirnov test \cite{hodges1958significance} on the average CAP scores for apolitical and divisive political discussions in Fig. \ref{fig:fig_ADvsDPD_cap_dist}, we see higher average CAP scores in divisive political discussions ($p < 0.001$).
Comparisons using other measures of influence in the retweet network---retweet PageRank and retweet h-index--- yields a similar result.
This is demonstrated by the violin plots in Fig. \ref{fig:fig_ADvsDPD_cap_dist}, which show the CAP distributions in divisive political and apolitical discussions for the top decile of users by each influence measure.
Note that while the Kolmogorov-Smirnov test indicates difference in the distributions at the significance level $p < 0.001$ for retweet count and PageRank, this difference is not statistically significant for the retweet h-index.
Nevertheless, 
this indicates that the participants in divisive political discussions come from a population that contains more inauthentic activity than participants in apolitical discussions, particularly among users that tend to be more influential in the retweet network with respect to PageRank centrality and retweet count.

\begin{figure}[!htb]
	\centering
	\captionsetup{width=.9\linewidth, format=hang}
	\includegraphics[width=.67\linewidth]{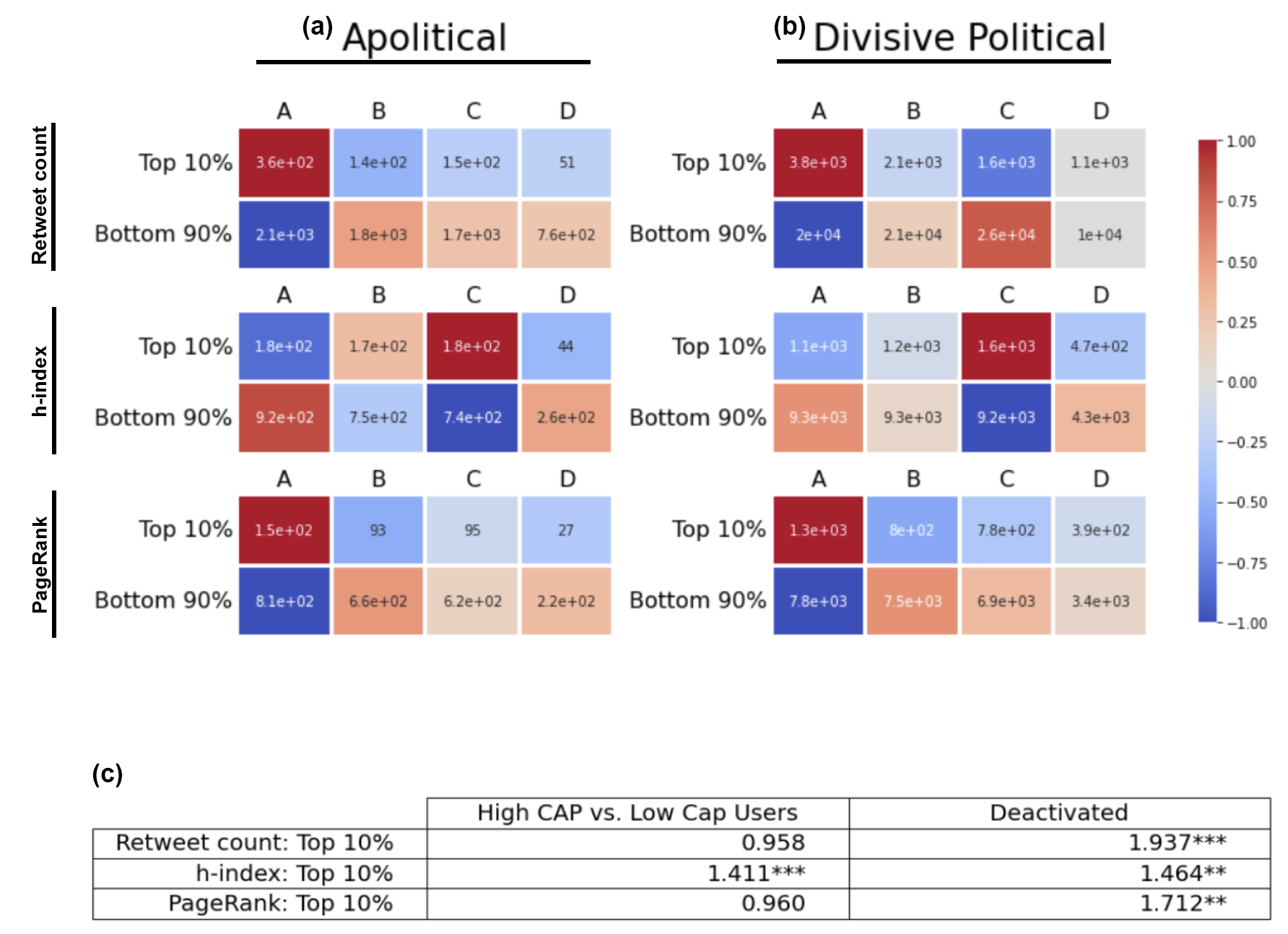}
	\caption{ \footnotesize 
	The rows in both the heat maps and the table correspond to retweet count, retweet h-index, and retweet PageRank, from the top row to the bottom row.
	\textbf{(a-b)}
	The heat maps show the enrichment-depletion patterns of users in the top-decile with respect to each measures, with the same vertical order as in the table, in \textbf{(a)} apolitical and \textbf{(b)} divisive political discussions. 
	Users groups A (high-CAP), B (mid-CAP), and C (low-CAP), correspond to the three types with respect to our account automation typology. 
	User group D indicates deactivated or suspended accounts.
	For each cell, the quantities are centered around the values with a randomly assigned distribution, conditional on the observed population of each row and column, and normalized to a $-1$ to $+1$ range.
	The annotations within each cell show the raw quantities.
	\textbf{(c)}
	The `Deactivated' column in the table shows the odds ratios of deactivation or suspension in divisive political discussions compared with apolitical discussions, for users belonging to the top decile by each of the three influence measures---retweet count, retweet h-index, and retweet PageRank.
	Similarly, the column labeled `High CAP vs. Low Cap Users' in the table shows the odds ratios of belonging to the high-CAP group of users (A), as opposed to the low-CAP group (C), for the same topics and with respect to the same influence metrics.
	Chi-square test is performed to verify statistical significance.
	`$*$', `$**$', and `$***$' indicate significance levels $0.1$, $0.05$, and $0.01$, respectively.
	}
	\captionsetup{labelformat=empty}
	\label{fig:fig_heatmap_influence}
\end{figure}

Beyond our relative measure of CAP scores, the deactivation or suspension of accounts shortly after their participation in a trending discussion provides further evidence of inauthentic behavior. 
Inauthentic accounts are much more likely to be suspended by Twitter or deactivated for suspicious activities than genuine Twitter users. 
We see higher levels of deactivation or suspension of accounts shortly after participation in divisive political discussions than apolitical discussions.
Specifically, we find that the odds ratio of suspension or deactivation of accounts in divisive political discussions to apolitical discussions
is approximately 1.06 (with $\chi^2$ statistic $> 4$ and $p<0.05$), which indicates that participants in divisive political discussions are more likely to be deactivated or suspended after participating in the discussion.
Performing the same analysis on only the top 10\% most retweeted accounts, we find that the odds ratio of suspension or deactivation in divisive political discussions to apolitical discussions is even larger, at 1.94 (with $\chi^2$ statistic $> 19$ and $p<0.001$). 
Similar patterns are observed for suspension or deactivation among users in the top decile by retweet PageRank and retweet h-index, with odds ratios of deactivation in divisive political to apolitical discussions 1.71 ($p<0.05$) and 1.47 ($p<0.05$), respectively.
Figure \ref{fig:fig_heatmap_influence} shows these odds ratios, as well as enrichment-depletion patterns for users grouped by activation status and user types with respect to our inauthenticity typology.
This indicates that there is even a stronger evidence of inauthentic activities in divisive political discussions if we focus on the most influential accounts in the discussion.

\subsection*{Participation in Trending Discussions} \zlabel{section_results_subsection_participationPatterns}

In addition to prevalence, Twitter data enables us to measure how inauthentic accounts behave during diverse types of trending topics. 
Discussions on Twitter are bursty, gaining traction, peaking, and re-equilibrating as users move on to the next topic or event. 
Typically, first movers begin an online conversation, a trending topic peaks within a few days, after which engagement slows and the topic stops trending. 
Drawing on our typology of authentic and inauthentic accounts, we can examine when different types of accounts entered the discussion of each topic, as well as how they behaved throughout the trending period. 
We construct the distribution of these patterns using time-series data of the tweets partitioned by user types.
Using the CAP scores from Botometer, we group users into three types---automated and semi-automated bots (group A), bot-assisted humans and trolls (group B), and genuine users (group C)--- falling within mutually exclusive CAP intervals, in decreasing order from group A to group C.
Our manual exploration of the data guides this large-scale systematic analysis of the temporal patterns and the user type configuration of the network corresponding to each discussion.
This includes the manual annotation of over 650 accounts with respect to features that characterize each type in our typology, as well as inspection of most influential (i.e. retweeted) users in each network cluster for each discussion. 
We use this manual annotation to confirm the expected order of CAP scores across user types according to our typology.
The manual annotation is subsequently matched against the CAP scores to find the interval boundaries for each type.
In particular, we find the boundaries on CAP intervals, such that the overall mismatch between manual annotations of the of accounts falling in the same group is minimized. 
The details of the criteria we use and the validation process are described in the \ztitleref{section_Methods_subsection_typology} subsection.

\begin{figure}[!htb]
	\centering
	\captionsetup{width=.9\linewidth, format=hang}
	\includegraphics[width=.63\linewidth]{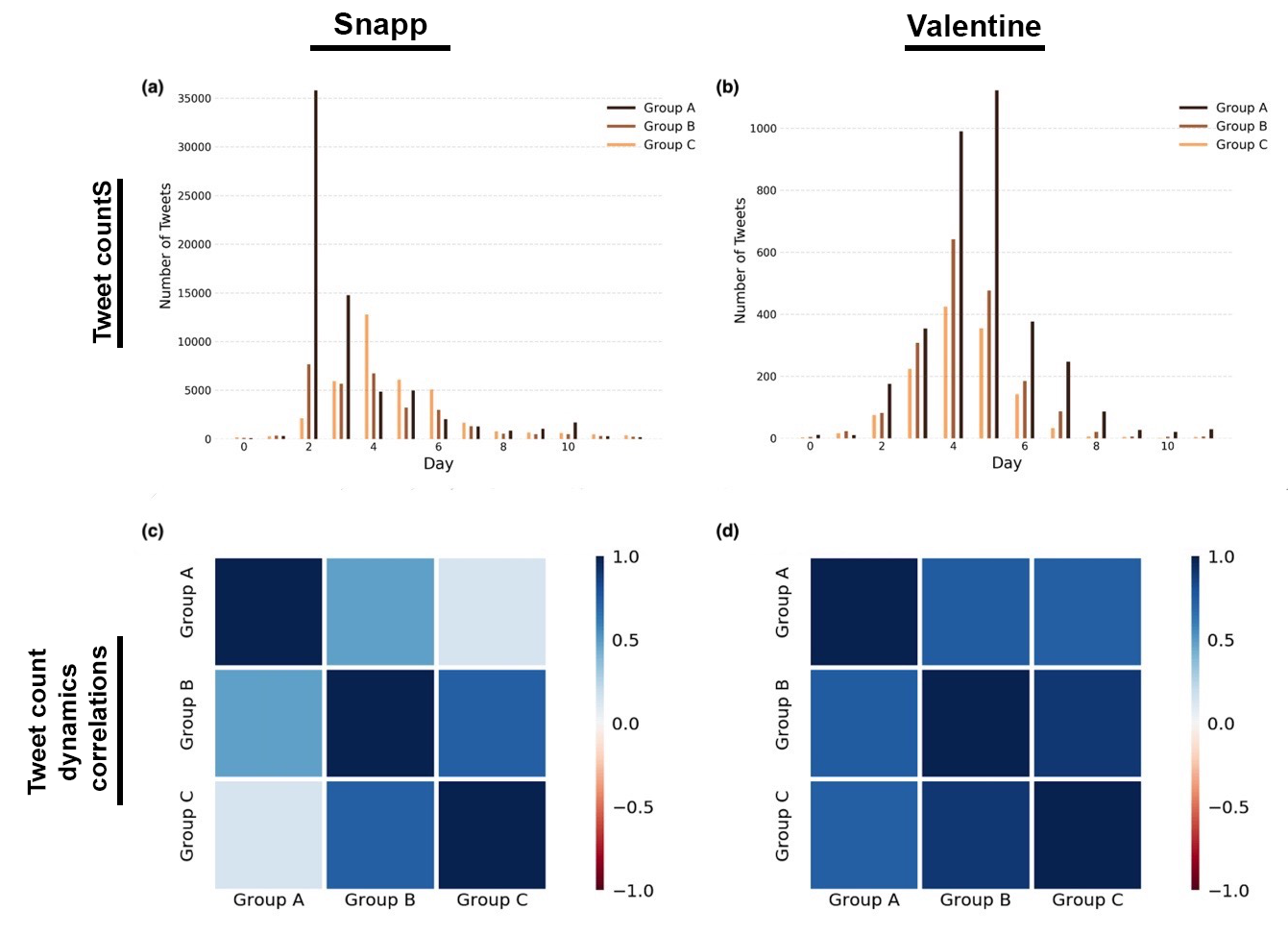}
	\caption{ \footnotesize Top row: The number of tweets by each group of users in each $24$-hour time interval since the beginning of the discussions.  (\textbf{a}) Snapp, a divisive political discussion. 
	(\textbf{b}) Valentine, an apolitical discussion.
	Bottom row: The Spearman correlation between the changes in the number of tweets from one day to the next for different groups of users.
	(\textbf{c}) Snapp, a divisive political discussion. 
	(\textbf{d}) Valentine, an apolitical discussion.
	The users are grouped, according to our typology described in the \ztitleref{section_Methods_subsection_typology} subsection, into automated inauthentic accounts (group A), bot assisted humans and trolls (group B), and genuine users (group C).
	The data for the plots in the left column (Snapp) contains 144347 tweets, and the data for the plots in the right column (Valentine) contains 7962 tweets.
	}
	\captionsetup{labelformat=empty}
	\label{fig:participationTrend_tweets_and_correlation}
\end{figure}

We find substantially different temporal behavior across different discussion types. {
To illustrate this dynamic, we provide evidence from one divisive political discussion about a female passenger who was refused service by a driver from Snapp, a popular ride-sharing service, who accused her of not complying with Iran's mandatory hijab law. Given existing controversies around the issue, crossing personal freedom and women's rights, the discussion became a trending divisive political discussion. We contrast this with one apolitical topic about Valentine's day. Although observance of Valentine's day is potentially controversial and could be associated with political or ideological orientations, this Twitter discussion remained apolitical throughout the period when it was trending. Results from all other apolitical and divisive political topics can be found in Appendix A. 
}
In divisive political discussions, we see that automated inauthentic accounts (group A) tweet earlier on and more often than bot assisted humans and trolls (group B) and genuine users (group C), helping to drive the topic to trend. 
This pattern is displayed in Fig. \ref{fig:participationTrend_tweets_and_correlation}, which shows the temporal dynamics of participation about the Snapp discussion. 
This stands in contrast to apolitical topics, where
all groups follow the same temporal tweeting pattern, as it is demonstrated for the Valentine discussion in Fig. \ref{fig:participationTrend_tweets_and_correlation}.
In order to quantitatively validate our observation, we compute the Spearman correlation coefficients for number of tweets per group for each discussion.
The results, visualized in Fig. \ref{fig:participationTrend_tweets_and_correlation}, reveal highly correlated temporal dynamics between different groups in apolitical discussions, while negligible correlations between participation pattern of inauthentic users and genuine users in divisive political discussions.
This is in agreement with the qualitative difference observed in the tweet counts in Fig. \ref{fig:participationTrend_tweets_and_correlation}.

\subsection*{Meso-scale Structure and Communities of Inauthentic Accounts}\zlabel{section_results_subsection_communityStructure}

\begin{figure}[!htb]
	\centering
	\captionsetup{width=.9\linewidth, format=hang}
	\includegraphics[width=.7\linewidth]{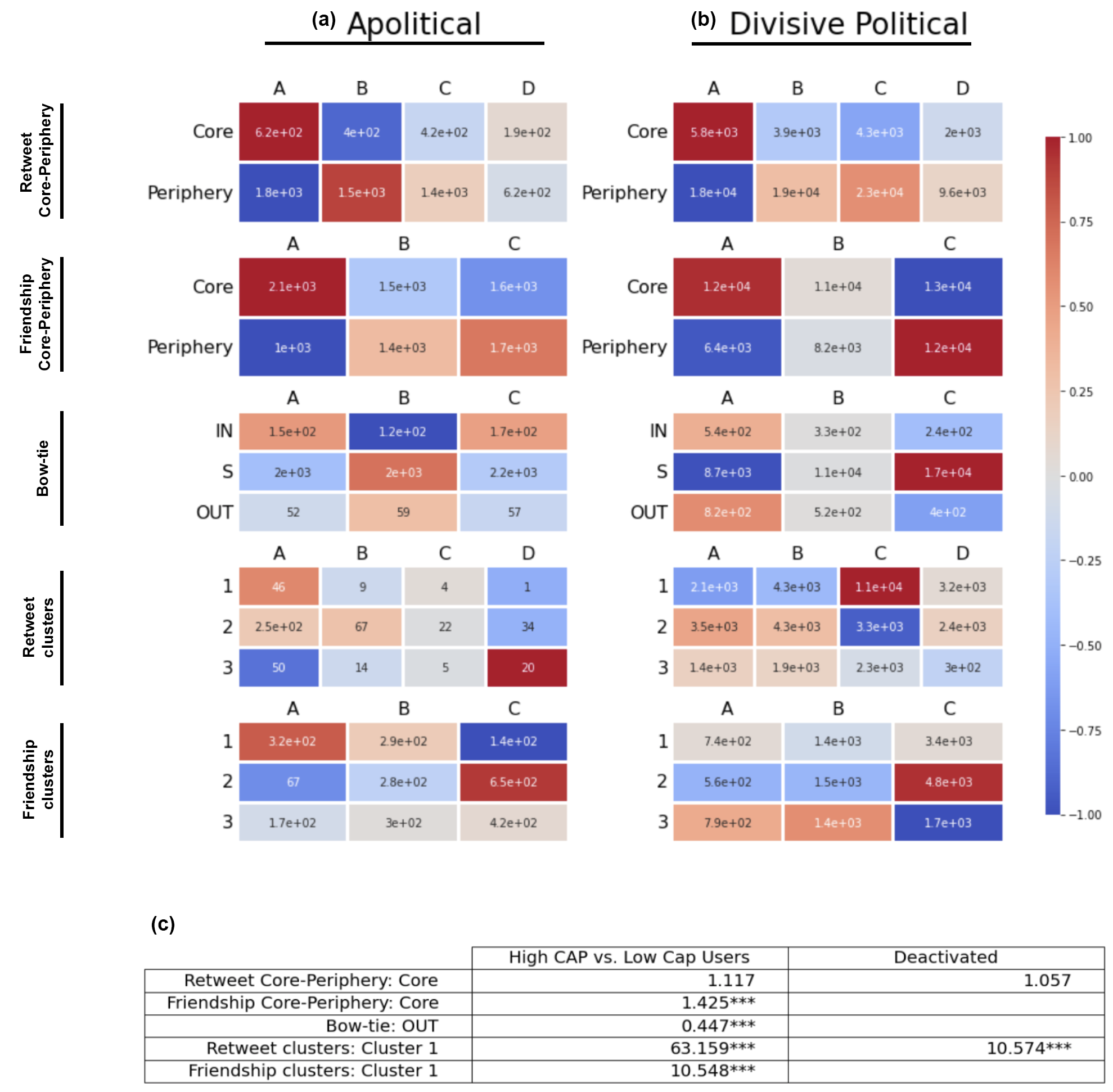}
	\caption{ \footnotesize 
	The rows in both the heat maps and the table correspond to retweet core-periphery, friendship core-periphery, friendship bow-tie, retweet clusters, and friendship clusters, from the top row to the bottom row.
	\textbf{(a-b)}
	The heat maps show the enrichment-depletion patterns of users in the top-decile with respect to each measures, with the same vertical order as in the table, in \textbf{(a)} apolitical and \textbf{(b)} divisive political discussions. 
	Users groups A (high-CAP), B (mid-CAP), and C (low-CAP), correspond to the three types with respect to our account automation typology. 
	User group D indicates deactivated or suspended accounts.
	For each cell, the quantities are centered around the values with a randomly assigned distribution, conditional on the observed population of each row and column, and normalized to a $-1$ to $+1$ range.
	The annotations within each cell show the raw quantities.
	Note that, since deactivation of an account disrupts the collection of neighbors of the account in the friendship network, nodes from group D are not considered in the heat maps corresponding to the friendship networks. 
	\textbf{(c)}
	The `Deactivated' column in the table shows the odds ratios of deactivation or suspension in divisive political discussions compared with apolitical discussions, for users belonging to each group of nodes in the retweet or friendship networks.
	That is, from the top row to the bottom row, core or periphery in retweet and friendship networks, bow-tie components in friendship networks, and 4 largest communities in retweet and friendship networks.
	Similarly, the column labeled `High CAP vs. Low Cap Users' in the table shows the odds ratios of belonging to the high-CAP group of users (A), as opposed to the low-CAP group (C), for the same topics and with respect to the same partitioning of the network.
	Chi-square test is performed to verify statistical significance.
	`$*$', `$**$', and `$***$' indicate significance levels $0.1$, $0.05$, and $0.01$, respectively.
	}
	\captionsetup{labelformat=empty}
	\label{fig:fig_heatmap_meso}
\end{figure}

Examining the structure of communities participating in trending discussions provides insight into the dynamics of inauthentic behavior. 
In each of the divisive political discussions we observe at least one major cluster with a higher aggregate level of inauthentic activity than that of the entire network. 
In apolitical discussions on the other hand, we see less variation in levels of inauthentic activity across all clusters. 
This is illustrated in Figure \ref{fig:fig_heatmap_meso}, which shows the enrichment patterns of each type of account with respect to our typology, and deactivated or suspended account, in each of the 4 largest network communities in the friendship and retweet networks.
The odds ratios of a user from the high-CAP group belonging to the largest network communities in divisive political discussions to that in apolitical discussions are substantially large and statistically significant with $p<0.001$ in the Chi-square test (see the table in Fig. \ref{fig:fig_heatmap_meso}).

Our analysis of the distribution of different user types across the retweet and friendship networks further reveals differences between divisive political and apolitical discussions in terms of the core-periphery network structure, which is related to amplification of exposure to content across the network on Twitter \cite{barbera2015critical}.
The results, visualized in Fig. \ref{fig:fig_heatmap_meso}, reveal that inauthentic groups of users tend to be significantly more concentrated in the core, in both friendship and retweet networks and across both types of discussions.
On the other hand, while in divisive political discussions low-CAP users are highly concentrated in the periphery and largely absent from the core of the friendship networks, the difference in apolitical networks is relatively less noticeable. 
There is no considerable difference in the enrichment of genuine users in core and periphery of the retweet networks.

The distribution of user types over the friendship networks also reveals significant differences between apolitical and divisive political discussions with respect to another meso-scale structural partitioning of the network---bow-tie structure \cite{broder2000graph}.
The bow-tie structure of directed networks identify where the nodes stand with respect to the direction of information flow, and are known to correspond to discursive communities in online social networks \cite{mattei2022bow}.
We consider the strongly connected core of the bow-tie (`S'), the immediate incoming gate of the S component (`IN'), and the immediate outgoing gate of the S component (`OUT') in the friendship network.
More details on how we detect and utilize the bow-tie structure can be found in the \ztitleref{section_Methods_subsection_networkAnalysis} subsection.
In the context of friendship networks of Twitter discussions, the components of the bow-tie mark the direction of exposure, or equivalently the potential propagation of content, among users engaged in the discussion.
Since the direction of the edges are from followers to friends, the OUT component could be viewed as the primary source of content propagation.
Our findings show relatively high concentration of high-CAP users in the OUT component of the bow-tie within the divisive political discussions, while this is not the case for apolitical discussions (see Fig. \ref{fig:fig_heatmap_meso}).
This signals relatively higher potential of high-CAP users for impacting the content of the discourse in divisive political discussions.
On the other hand, high-CAP users are largely absent from the S component of the bow-tie in divisive political discussions, which indicates that genuine users have higher potentials for facilitating the circulation of content in these discussions.

\subsection*{Differences in Content}\zlabel{section_results_subsection_content}

\begin{figure}[!htb]
	\centering
	\captionsetup{width=.9\linewidth, format=hang}
	\includegraphics[width=.65\linewidth]{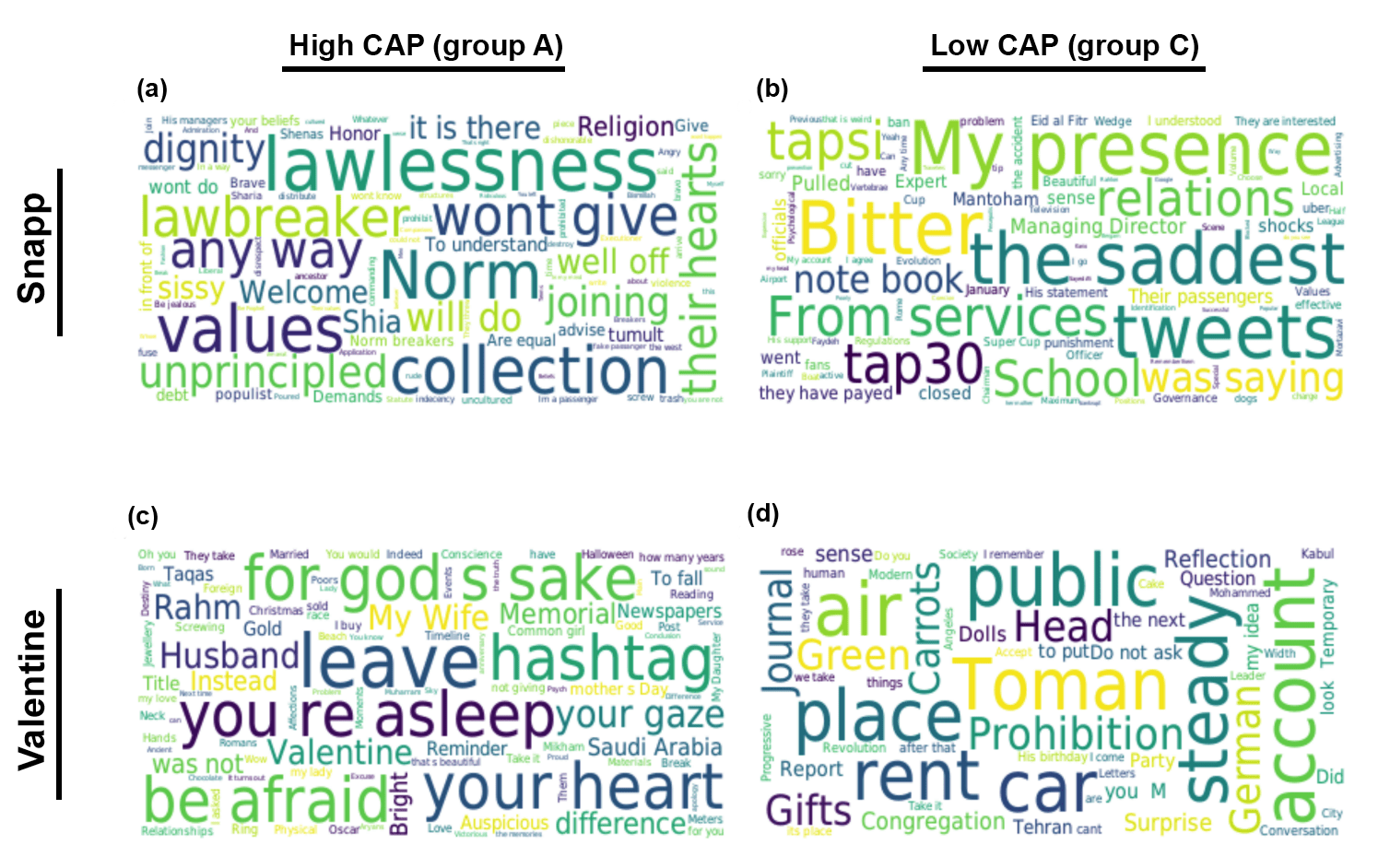}
	\caption{ \footnotesize  Difference word clouds by CAP for two discussions. 
	(\textbf{a-b}) Snapp, a divisive political discussion. 
	(\textbf{c-d}) Valentine, an apolitical discussion.
	(\textbf{a}) and (\textbf{c}) include the top 5\% most frequent words across the high-CAP group which are absent from the top 10\% most frequent words used by the low-CAP group. 
	(\textbf{b}) and (\textbf{d}) include the top 5\% most frequent words across the low-CAP group which are absent from the top 10\% most frequent words used by the high-CAP group.}
	\captionsetup{labelformat=empty}
	\label{fig:wordclouds_AvsC_valentineANDsnapp}
\end{figure}

\begin{figure}[!htb]
	\centering
	\captionsetup{width=.9\linewidth, format=hang}
	\includegraphics[width=.83\linewidth]{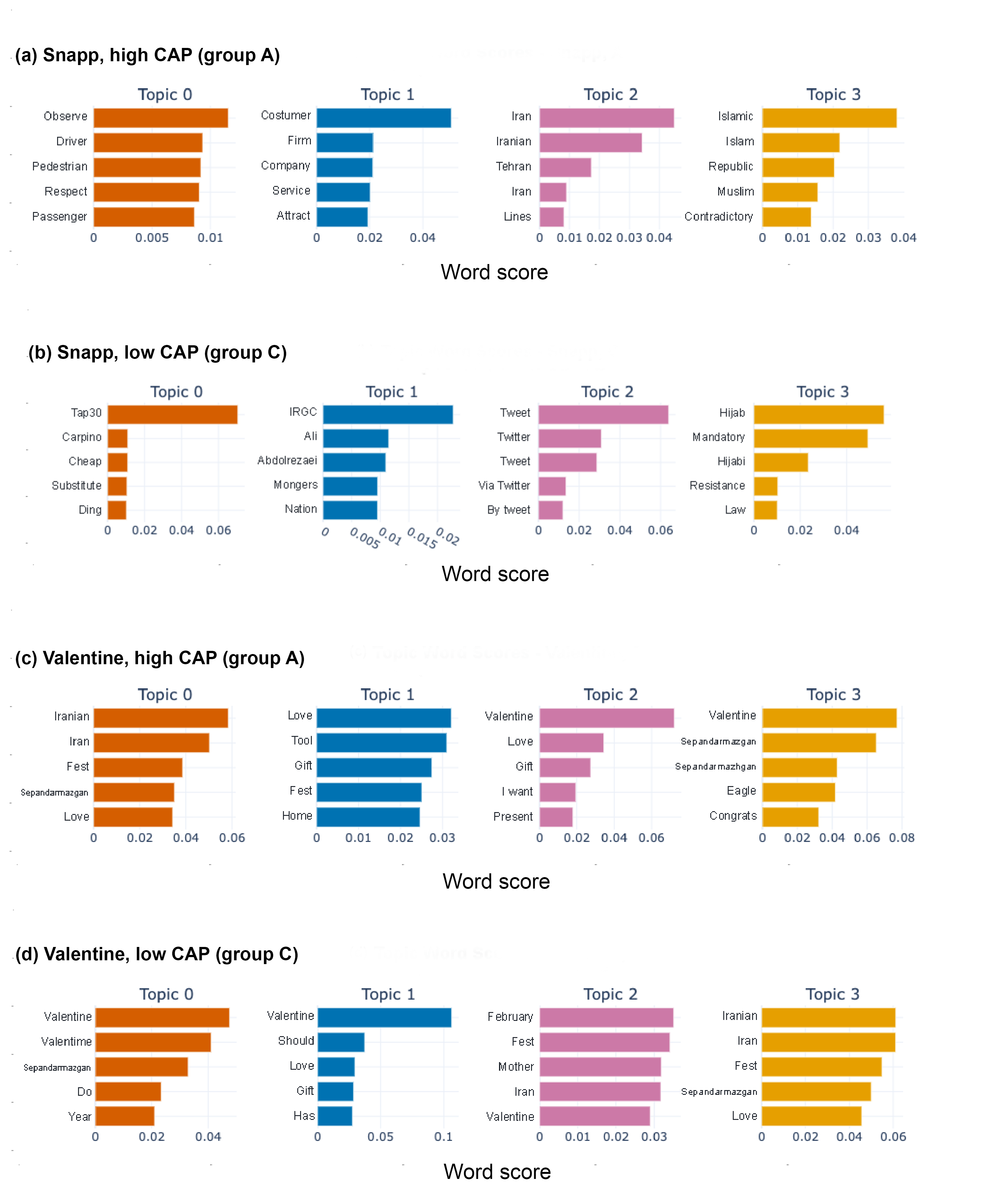}
	\caption{
	\footnotesize  Salient topics and highest scored words within each topic, obtained from BERTopic, for tweets from \textbf{(b,d)} low-CAP (C) and \textbf{(a,c)} high-CAP (A) user types in two discussions.
	(\textbf{a-b}) Snapp, a divisive political discussion. 
	(\textbf{c-d}) Valentine, an apolitical discussion.}
	\captionsetup{labelformat=empty}
	\label{fig:topics_AvsC_valentineANDsnapp}
\end{figure}

Comparing the content of tweets produced by communities with high and low levels of inauthentic activity, in divisive political discussions, we notice a more polarizing language among communities with higher levels of inauthentic activity. 
However, when we look at apolitical discussions, we see little difference in the language used by communities with higher and lower levels of inauthentic activity. 
This can be seen from Fig. \ref{fig:wordclouds_AvsC_valentineANDsnapp} and Fig. \ref{fig:topics_AvsC_valentineANDsnapp}, which, respectively, display the most frequent words and salient topics, among users grouped by our inauthenticity typology.
The word clouds in Fig. \ref{fig:wordclouds_AvsC_valentineANDsnapp} compare the highest frequency words (translated from Farsi to English) in communities of genuine users (group C) to those produced by automated and semi-automated bots (group A), in the divisive political discussion about Snapp and the apolitical discussion about Valentine. 
More specifically, they show the words that are among the top 5\% most frequent words in the tweets posted by automated inauthentic users, but not among the top 10\% most frequent words used by genuine users.
While here we discuss the Snapp and Valentine discussions in detail, the main observations in these word clouds generally hold across other discussions of the same type, i.e. divisive political and apolitical discussions.
We further use an unsupervised transformer-based topic detection algorithm to find salient topics along with most representative words within each topic for each group of users within each discussion.
Additional details about the content analysis are explained in the \ztitleref{section_Methods_subsection_contentAnalysis} subsection.
Our observation from comparing top words within the first 4 topics (Fig. \ref{fig:topics_AvsC_valentineANDsnapp}) is aligned with what is observed in the word clouds, confirming the polarizing language within each group in divisive political discussions.

In order to interpret the observations from this  content analysis, we provide further context on the Snapp discussion. The Snapp discussion occurred after a driver employed by Snapp, the country's biggest ride-sharing enterprise, refused to take a passenger to her destination because she had not been observing Hijab, which is mandatory according to a rather controversial clothing law for women in Iran. The topic started to trend on Twitter after the passenger tweeted about the incident, criticizing the driver for making her leave the car, and demanding accountability from the company, Snapp. Subsequently, a dispute erupted between Twitter users expressing a range of opinions from supporting the passenger to reproaching her. 
In Fig. \ref{fig:wordclouds_AvsC_valentineANDsnapp} we observe that inauthentic users (group A)  frequently used polarizing words ideologically aligned with the Iranian state.
For example, the words `Law-breaker`, `Lawlessness`, and `Unprincipled` carry a critical tone towards the clothing of the passenger involved in the incident. 
Moreover, the words `Value`, `Norm`, and `Dignity` are also representative of the vocabulary corresponding to the same ideology. 
However, genuine users (group C) more frequently used words such as `Bitter`, `Punishment`, and `the saddest`, signifying their sympathetic stance with the passenger in the story. 
Similarly, the words with highest word scores among the top-4 detected topics show qualitative differences between the content of tweets from inauthentic and genuine users, shown in Fig. \ref{fig:topics_AvsC_valentineANDsnapp}.
For instance, `Observe', `Respect', `Muslim', representative words of topics discussed by high-CAP group, imply advocacy in favor of the driver involved in the event, which is in line with the position implied by the highly-frequent words in the corresponding word cloud.
Meanwhile, words such as `IRGC', `Tap30', and `Mandatory', which are representative of salient topics in the content produced by the low-CAP accounts, support the opposite stance.

By contrast, the other pair of word clouds in Fig. \ref{fig:wordclouds_AvsC_valentineANDsnapp}, which display the characteristic words across authentic and inauthentic users in the apolitical discussion about Valentine's day, do not display meaningful differences between the two groups.   
Neither the words appearing as highly frequent words only among the inauthentic users (group A) nor those frequently used only by the genuine users (group C), convey any particular meaning or otherwise signal a divide in the content of discussion between the low-CAP and high-CAP users.
This also holds for the representative words of the topics detected through topic modeling on the tweets in the Valentine discussion, where we do not observe a substantive difference between the topics corresponding to the two groups of users (see Fig. \ref{fig:topics_AvsC_valentineANDsnapp}).

\begin{figure}[!htb]
	\centering
	\captionsetup{width=.9\linewidth, format=hang}
	\includegraphics[width=.74\linewidth]{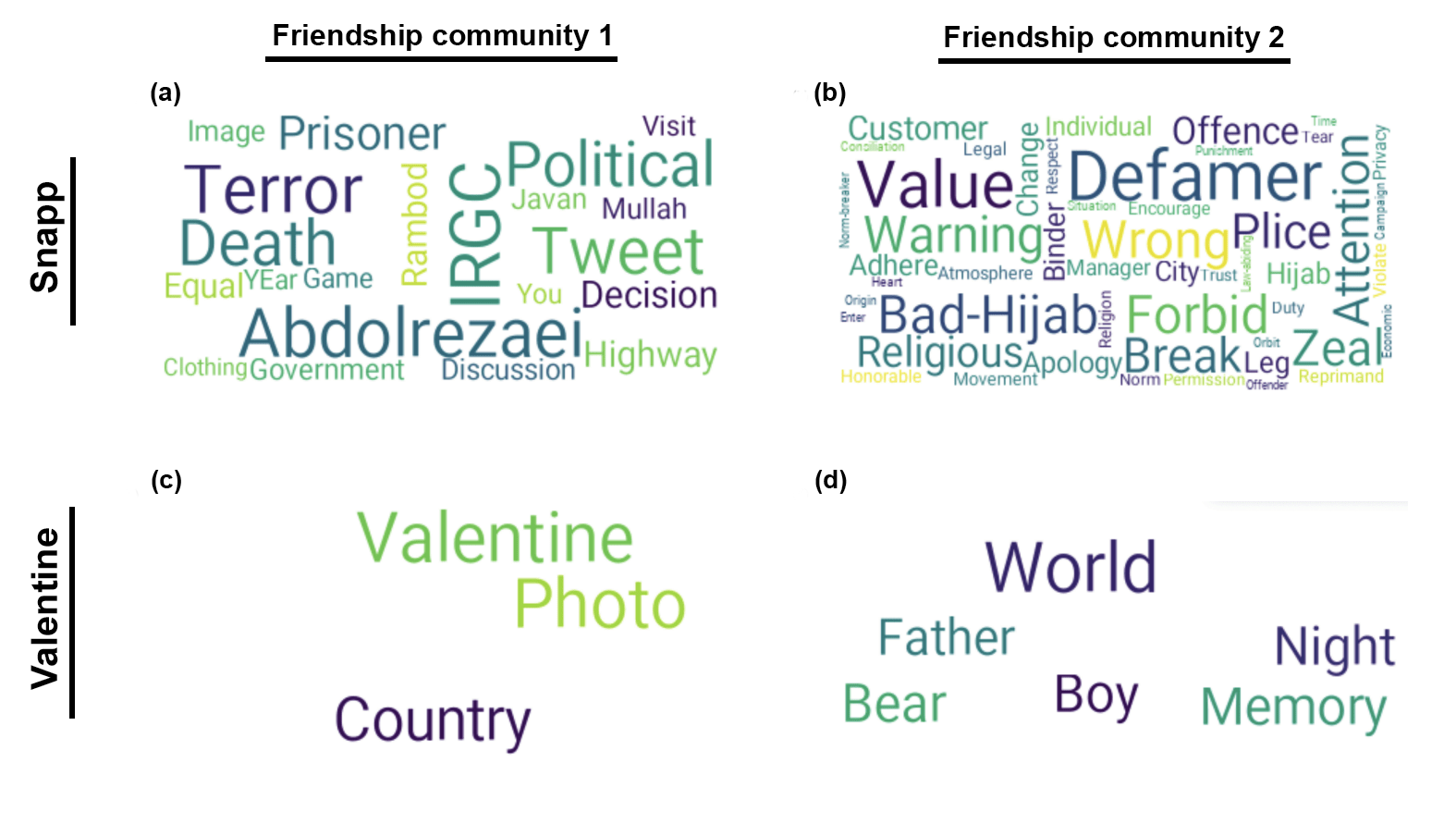}
	\caption{ \footnotesize  Difference word clouds for friendship network communities corresponding to two discussions. 
	(\textbf{a-b}) Snapp, a divisive political discussion.
	(\textbf{c-d}) Valentine, an apolitical discussion.
	(\textbf{a}) and (\textbf{c}) include the top 5\% most frequent words used by users in community 1 which are not among the top 10\% most frequent words used by users in community 2.
	(\textbf{b}) and (\textbf{d}) include the top 5\% most frequent words used by users in community 2 which are not among the top 10\% most frequent words used by users in community 1.}
	\label{fig:wordclouds_community_valentineANDsnapp}
\end{figure}

\begin{figure}[!htb]
	\centering
	\captionsetup{width=.9\linewidth, format=hang}
	\includegraphics[width=.83\linewidth]{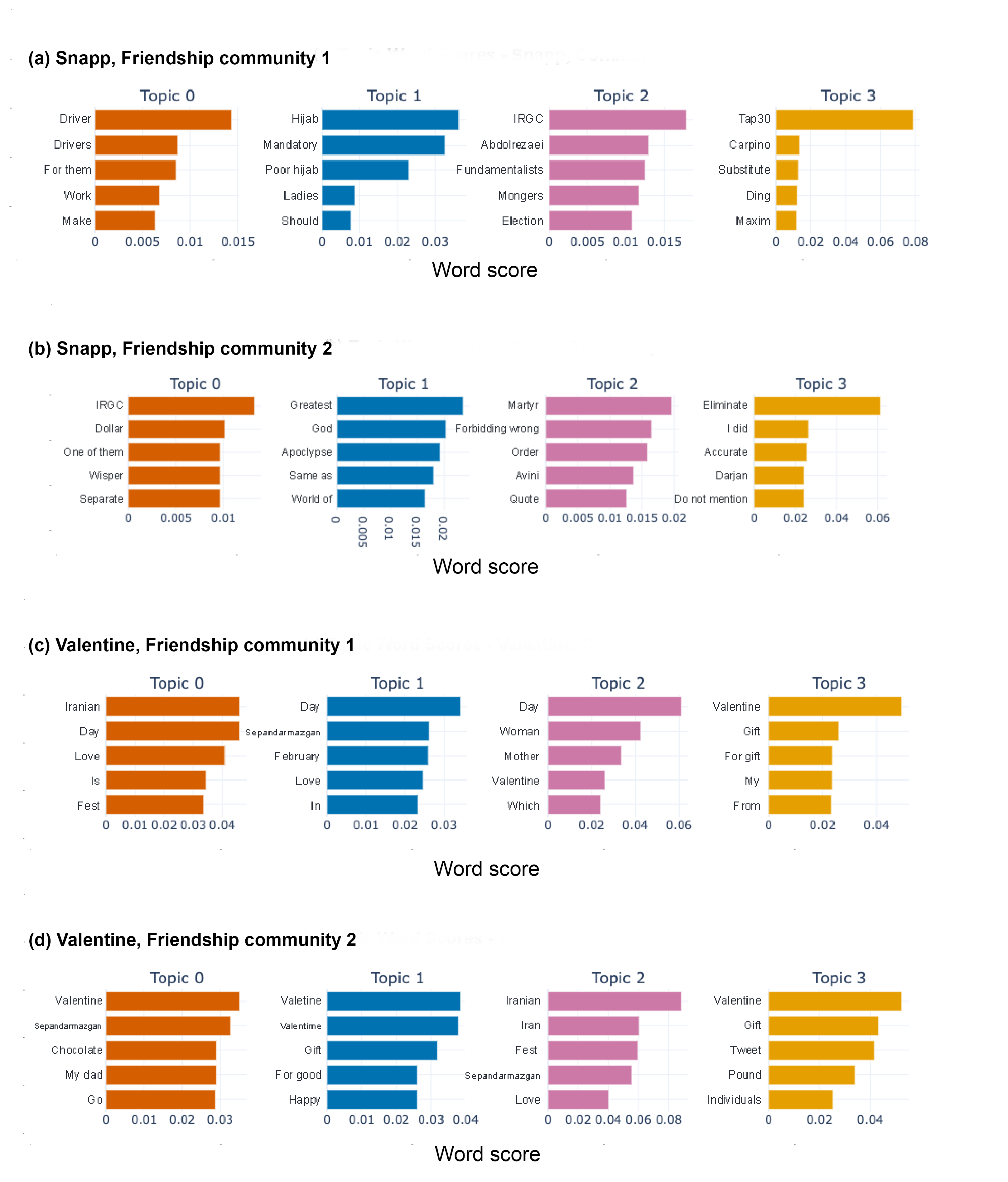}
	\caption{ \footnotesize  Salient topics and highest scored words  within each topic, obtained from BERTopic, for tweets from users in two friendship communities.
	(\textbf{a-b}) Snapp, a divisive political discussion. 
	(\textbf{c-d}) Valentine, an apolitical discussion.}
	\captionsetup{labelformat=empty}
	\label{fig:topics_community_valentineANDsnapp}
\end{figure}

In addition to comparing the content of tweets from authentic and inauthentic users (as measured by their CAP scores), we make the same comparison for users belonging to different clusters in friendship networks of each trending discussion.
Fig. \ref{fig:wordclouds_community_valentineANDsnapp} displays the difference between the high frequency words used by users belonging to two major communities in the friendship network of the Snapp discussion, which is a divisive political discussion.
This is shown in word clouds constructed in the same fashion as the word clouds in Fig. \ref{fig:wordclouds_AvsC_valentineANDsnapp}, i.e. each word cloud is showing the words that are among the top 5\% most frequent words in one community but not among the top 10\% in the other.
The words in the word cloud on the left clearly signal a conservative rhetoric in the corresponding tweets, while the word cloud on the right includes words from the common vocabulary of the dissidents and the groups that are critical of the status quo.
Similar to the comparison between the content from low-CAP and high-CAP users, we also compare the detected topics and their representative words in each of the two friendship communities.
This comparison, shown in Fig. \ref{fig:topics_community_valentineANDsnapp}, provides further evidence for polarization of content between two major communities in the friendship network of the Snapp discussion.

Note the significance of this observation is that this difference is observed between the communities in the friendship network.
Friendship networks are structural constructs formed by follower-friend connections between the participants in the discussion.
These connections are often made prior to the discussion, and have no direct link to the content of the tweets.
Meanwhile, the network clustering algorithm is also completely blind to the text of the tweets or otherwise-content-related features.
Yet we can observe a clear difference between the content of tweets tweeted by users from different network clusters, which is indicative of the political orientation of these users.
Due to how user activities appear in their followers' feed, a community in the friendship network approximately corresponds to a neighborhood of users with significantly overlapping exposure to content. 
Hence, the clear difference in the content of the tweets tweeted by users from different network communities speaks to the differences in the exposure of these users.
In other words, users belonging to different communities are exposed to qualitatively different contents, with opposing political rhetoric, which is a sign of presence of echo chambers in divisive political discussions.
Notice that there is no meaningful difference between {the detected topics and} highly frequent words used by different friendship communities in apolitical discussions, as it is shown in Fig. \ref{fig:wordclouds_community_valentineANDsnapp} and Fig. \ref{fig:topics_community_valentineANDsnapp}.

\subsection*{Inauthentic Activity in Echo Chambers}\zlabel{section_results_subsection_echochambers}

\begin{figure}[!htb]
	\centering
	\captionsetup{width=.9\linewidth, format=hang}
	\includegraphics[width=.6\linewidth]{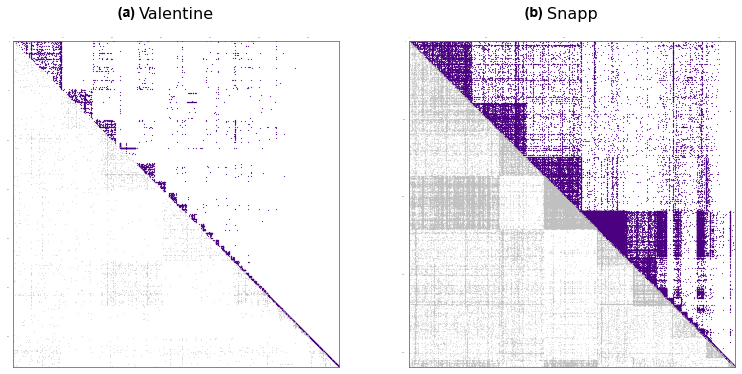}
	\caption{
	\footnotesize The symmetrized retweet networks for the Twitter discussions \textbf{(a)} Valentine and \textbf{(b)} Snapp.
	In the upper triangle the rows and columns are reordered such to group nodes by their retweet communities.
	In the lower triangle, the reordering groups nodes by the community they belong to in the corresponding friendship network.
	Excluding the isolated nodes, the retweet network of the Snapp discussion has 21029 nodes and 69565 edges, and the retweet network of the Valentine discussion has 3705 nodes and 4491 edges.}
	\captionsetup{labelformat=empty}
	\label{fig:RtNw_with_RtComm_vs_FrComm}
\end{figure}

Our comparison of the community structures in the retweet and friendship networks provides more insight into the communities in which inauthentic accounts insert themselves. 
This analysis reveals that echo chambers are much more prevalent in divisive political conversations relative to apolitical conversations. 
Figure \ref{fig:RtNw_with_RtComm_vs_FrComm} compares the communities in the retweet and friendship networks of the Snapp (divisive political) and the Valentine (apolitical) discussions.
The upper triangle in each plot on Fig. \ref{fig:RtNw_with_RtComm_vs_FrComm} shows the symmetrized adjacency matrix of the retweet network with a permutation of rows and columns that groups the nodes by their retweet community.
The lower triangle on the other hand, shows the same symmetrized adjacency matrix but with an alternative permutation which groups nodes by the friendship communities they belong to. 
As the lower triangle of the left plot in Fig. \ref{fig:RtNw_with_RtComm_vs_FrComm} suggests, the friendship communities of the Snapp discussion induce a clustered structure on the retweet network.
Furthermore, comparing the lower and upper triangles, we can see that this induced clustering has a rather considerable overlap with the retweet communities. 
This, however, is not the case for the Valentine discussion, as we can see from the plot on the right of Fig. \ref{fig:RtNw_with_RtComm_vs_FrComm}, which does not show a qualitatively meaningful clustering of the retweet network induced by the friendship communities.

Note that a retweet community is the group of users among which circulation of content takes place, while a friendship community signifies the neighborhood in the network where users are exposed to a mutual content.
The overlap between the two communities, which we see in the Snapp discussion in Fig. \ref{fig:RtNw_with_RtComm_vs_FrComm},  suggests the presence of echo chambers.
We see a similar pattern across divisive political discussions, indicating the presence of echo chambers, which we do not observe in apolitical discussions, where the friendship communities do not induce a pronounced clustering on the retweet network. 
This can be seen for the Valentine discussion, as an example of an apolitical discussion, in Fig. \ref{fig:RtNw_with_RtComm_vs_FrComm}.
As the sparsity pattern of the adjacency matrix of the retweet network ordered by the friendship communities shows (lower triangle), the firendship network communities are scattered across the retweet network of the Valentine discussion.
This is consistent across other apolitical discussions as well, which indicates lack of evidence for formation of echo chambers in this category of discussions.

\begin{figure}[!htb]
	\centering
	\captionsetup{width=.9\linewidth, format=hang}
	\includegraphics[width=.6\linewidth]{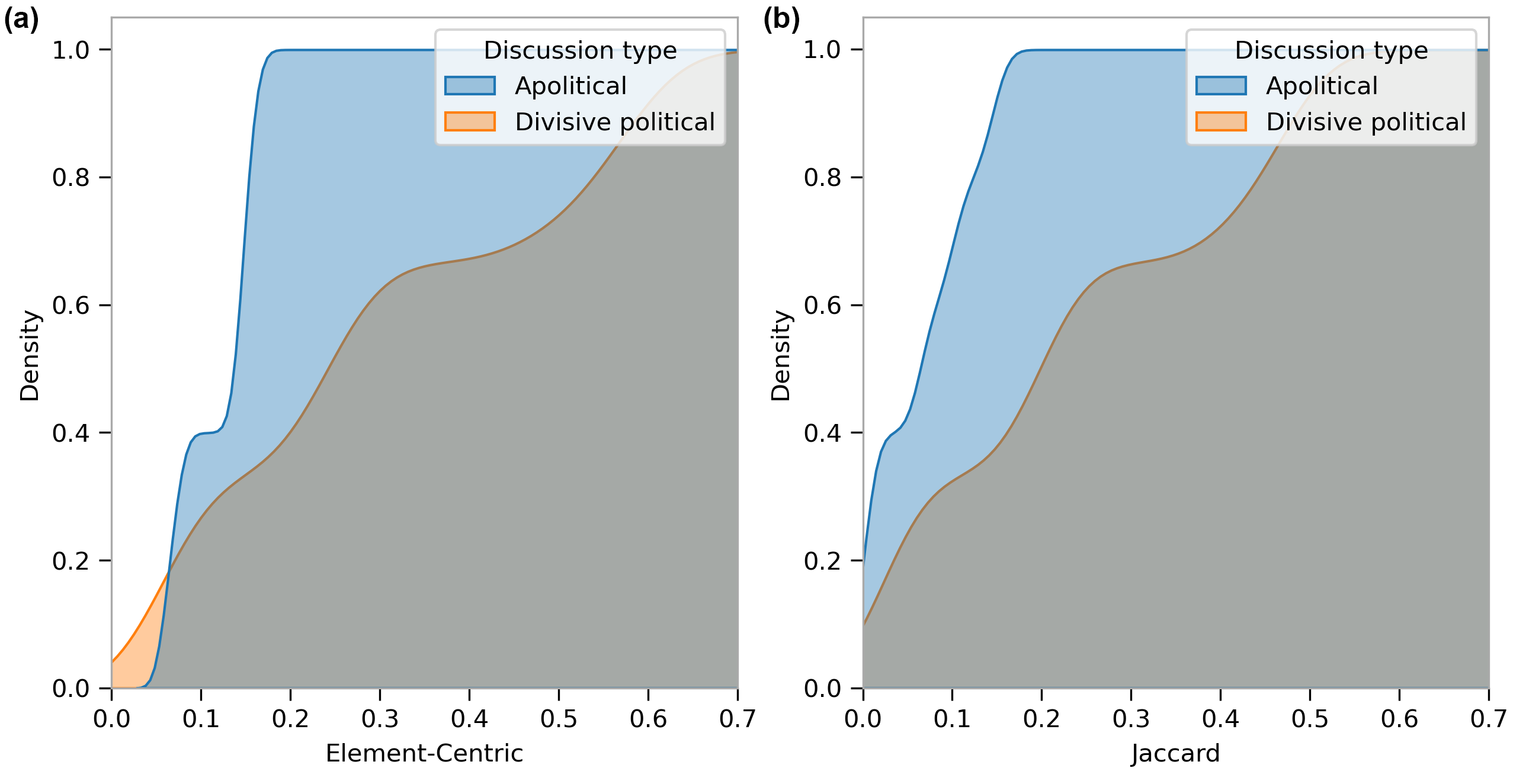}
	\caption{ \footnotesize The empirical cumulative distribution functions (CDF) of the similarities between the retweet and friendship networks of the apolitical (blue) and divisive political (orange) discussions, over 20 different runs of Louvain clustering for each network.
	That is, for each discussion, the ensemble contains 400 pairs of similarity values.
	Note that the diagrams use a Kernel Density Estimate to obtain smoothed distributions. 
	\textbf{(a)} The similarities are computed according to element-centric similarity \cite{gates2018element}.
	\textbf{(b)} The similarities are computed according to Jaccard index.}
	\captionsetup{labelformat=empty}
	\label{fig:sims_RtComm_vs_FrComm}
\end{figure}

Figure \ref{fig:RtNw_with_RtComm_vs_FrComm} visualizes the difference between divisive political and apolitical discussions in terms of the overlap between retweet and friendship communities in the case of two example discussions.
In order to further verify this result, we quantify the overlap between retweet and friendship communities, as well as polarization of the network structure, for each type of discussion. 
Given the randomness involved in the community detection method and the sensitivity of measures of similarities between two clusterings (see \cite{gates2018element}), instead of comparing only one pair of clustering outcomes, we consider the distribution of pairwise clustering similarities between retweet and friendship networks in an ensemble of network clusterings.
Using two measures of clustering similarities, Fig. \ref{fig:sims_RtComm_vs_FrComm} compares the empirical cumulative distribution functions (CDF) of the similarities between the retweet and friendship networks in apolitical and divisive political discussions.
The details about the similarity measures and how the empirical CDF's are observed are explained in the \ztitleref{section_Methods_subsection_networkAnalysis} subsection.
As we can see in Fig. \ref{fig:sims_RtComm_vs_FrComm}, the empirical CDF corresponding to divisive political discussions falls below that of the apolitical discussions across almost the entire range of similarity values, according to both measures.
Furthermore, the empirical CDF's corresponding to apolitical discussions reach $1$ very rapidly, at very small similarity values.
This means the retweet and friendship networks of apolitical discussions tend to be consistently dissimilar across several runs of the clustering algorithm, while the structure of friendship and retweet networks of divisive political discussions can lead the clustering algorithm to output considerably overlapping friendship and retweet communities.
In other words, in the light of the discussion above about the qualitative connection between echo-chambers and overlapping friendship and retweet communities, the observations in Fig. \ref{fig:sims_RtComm_vs_FrComm} confirm that the network structure of divisive political discussions is conducive to formation of echo-chambers, while this is not the case in apolitical discussions.

\begin{figure}[!htb]
	\centering
	\captionsetup{width=.9\linewidth, format=hang}
	\includegraphics[width=.58\linewidth]{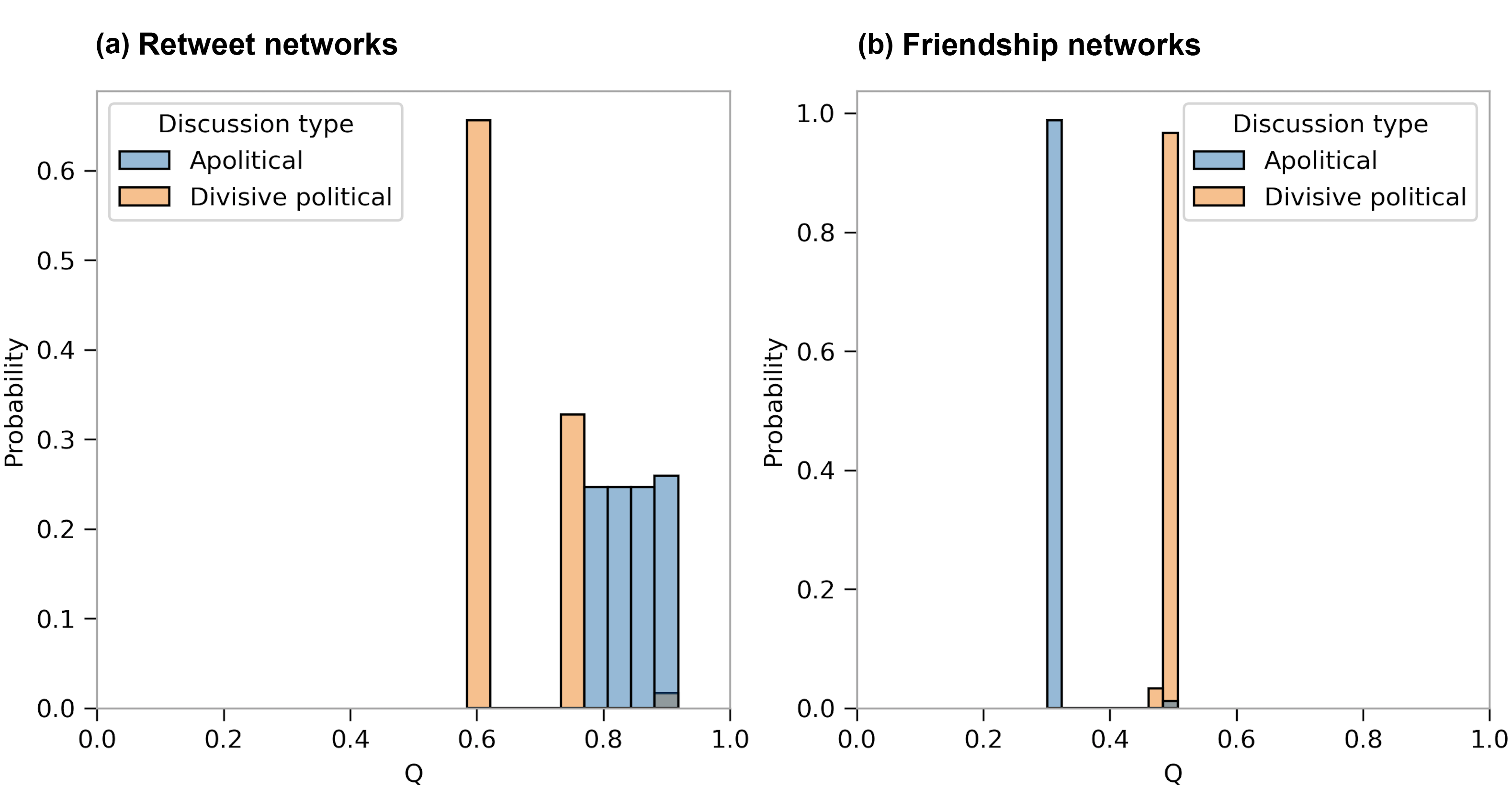}
	\caption{ \footnotesize The distribution of network modularities (denoted $Q$ on the horizontal axis) in the \textbf{(a)} retweet and \textbf{(b)} friendship networks of the apolitical (blue) and divisive political (orange) discussions, over 100 different runs of Louvain clustering for each network.}
	\captionsetup{labelformat=empty}
	\label{fig:Qs_Rt_and_Fr}
\end{figure}

As an additional assessment of structural polarization of the network, we compute the modularities of both types of networks for each discussion, the distributions of which, over ensembles of multiple runs of the clustering algorithm, are shown in Fig. \ref{fig:Qs_Rt_and_Fr}.
A large modularity means users are densely connected within each community, with relatively few out-community connections \cite{newman2006modularity}.
Hence, modularity is a measure of segregation of users into clusters, and is used to evaluate the potential for structural polarization in online social networks (see e.g. \cite{Conover_politicalPolarizationTwitter, livne2011party}).
We compute the modularity for 100 runs of the clustering algorithm on each network, which yields a distribution of modularity values for each network and discussion type.
The details are described in the \ztitleref{section_Methods_subsection_networkAnalysis} subsection.
These distributions are shown by the histograms in Fig. \ref{fig:Qs_Rt_and_Fr}, where we can see that the friendship networks in divisive political discussions tend to have relatively large modularity values concentrated over a narrow range, while this is not the case for apolitical discussions, where there seems to be a lack of inherently segregated network structure.
While the modularity values for the retweet networks in divisive political discussions have a narrower range, and those for apolitical discussions tend to have larger values, given what the retweet network represents, this does not convey information on echo chambers on its own.
However, in light of our observation on the difference in friendship modularities, the relative similarity in retweet modularities suggests that, while users retweet each other in a rather modular and segregated fashion in both types of discussion, the friendship structure of divisive political discussions makes them more prone to segregated exposure to the retweets.
This provides further evidence for conduciveness of the network structure to polarization and emergence of echo chambers in divisive political discussions, which is consistent with our interpretation of the overlap between friendship and retweet communities.

\section*{Discussion}\zlabel{section_Discussion}

Taken together, our analysis of inauthentic activity across trending topics in the Farsi Twittersphere demonstrate that inauthentic accounts are very active in divisive political discussions, where they often initiate trending topics and use polarizing language to advance partisan agendas. This stands in contrast to their activity in apolitical discussions, where inauthentic accounts are less prevalent and engage in similar discourse to genuine users. The network structure of these conversations provides insight into the communities in which inauthentic accounts operate. Divisive political conversations occur within echo chamber environments, where retweet networks and friendship networks overlap to a high degree. By contrast, apolitical conversations bridge the partisan divide, reaching users of diverse ideological persuasions. Inauthentic accounts are therefore able to advance partisan narratives in echo chamber environments where they can target specific partisan audiences.  

By analyzing the dynamics of inauthentic activity in the Iranian Twittersphere, our study contributes to a growing body of literature exploring online influence operations in several ways. First, we provide a typology of accounts enabling us to distinguish between automated inauthentic accounts, human-controlled inauthentic accounts, and genuine users. Second, we provide evidence examining the content, structure, and temporal dynamics of inauthentic activity, providing a rich multi-method characterization of inauthentic online behavior. Third, we provide evidence from the Farsi Twittersphere, an understudied context where inauthentic activity is quite prevalent. Fourth, unlike most existing studies, we cover diverse issue areas, enabling us to compare inauthentic behavior across divisive political, non-divisive political, and apolitical topics in the same analysis. 
The multifaceted analysis of several Twitter discussion spaces, allowed us to overcome common challenges posed by errors in bot-detection methods, missing data points, and method-induced ambiguities in isolated analysis techniques.
As a result, we are able to provide systematic descriptive evidence on how inauthentic accounts engage in online discourse on Farsi Twitter.

Despite these contributions, our study has several limitations that we hope future research will help to overcome. Our study is limited to just one platform---Twitter---and therefore does not allow us to characterize online inauthentic activity more generally.
Having said that, existing research on inauthentic accounts and opinion manipulators on Facebook, reveals partial similarities between features of these accounts and those identified in our typology for inauthentic Twitter accounts.
For instance, inauthentic accounts carrying information operations on Facebook are deployed for content creation and false amplification of content \cite{weedon2017information}.
Additionally, inauthentic accounts on both Facebook and Twitter have been used for engagement with authentic accounts for manipulating engagement statistics \cite{schliebs2021china}.
For detection of inauthentic activities on Facebook, combining manual and automatic labeling to improve reliability, previous works have found similar validation techniques to what is used in our study to be effective \cite{gupta2017towards}.
The authors have also pointed to the potential utility of network properties in improving detection methods on Facebook \cite{gupta2017towards}, and our study can point to directions for guided investigations of network features that could be most helpful to consider, given our findings regarding the patterns of inauthentic activities across network neighborhoods. We hope that future research will include cross-platform analyses that enable us to better understand the broader ecosystem of online coordinated inauthentic activity.
Additionally, measuring inauthentic activity is very challenging. While human validation of our measurement approach suggests it performs well at an aggregate level, existing automated approaches prevent us from accurately classifying accounts at the individual level, an important step for future research. 

Recent research demonstrates that state and non-state actors are increasingly leveraging social media platforms to run influence operations. Understanding how these actors operate across diverse issue areas and global contexts is therefore crucial from a policy perspective. We hope that online platforms will continue to make data on information operations available to researchers, to help us to improve our characterization and measurement of online inauthentic activity. In the absence of better data and measurement tools, we encourage researchers to draw on similar methodological approaches to those we present here to continue to characterize inauthentic activity across different issue areas, time periods, and contexts.

\section*{Methods}\zlabel{section_Methods_and_Materials}

In this section, we describe our topics of discussion, clarify our dataset and details of our data collection, and explain our data analysis methods. 
Moreover, in order to understand the configuration of our discussion spaces in terms of user types with respect to automation and authenticity, we introduce a typology which helps reduce ambiguities in our observations.

\subsection*{Selected Topics}\zlabel{section_Methods_subsection_topics}

Different topics of discussions in Twitter can shape discussion spaces with different qualities.
For instance, Smith et al \cite{Smith_polarizedCrowds} introduce six different types of political discussions in Twitter with respect to their social structure.
Our exploratory analysis of trending discussions in Farsi Twitter revealed three main types of topics of discussion with qualitative differences in the development of their trend, network structure, and participation of users.
In order to verify consistency in our observations across various discussions of each type, for the targeted phase of our data collection, we selected an overall of eleven topics including four apolitical discussions, four non-divisive political discussions, and three divisive political discussions.
The analysis in the \ztitleref{section_Results} section focuses on the characteristics of divisive political discussions compared against apolitical discussions.
Please note that the border between non-divisive and divisive political discussions is rather blur, and there is a transition zone between these two types.
While the contrast between divisive political and apolitical discussions is more evident in our results, allowing us to more clearly characterize these two categories of discussions, non-divisive political discussions mostly show mixed features.
The discussion topics are described in detail in Appendix B.

In order to make sure every discussion space has the potential to show certain characterizing features such as polarization or formation of echo chambers, all chosen subjects cross socio-political cleavages in the Iranian society. 
Therefore, although not directly political, all events can turn into a political handle for supporters of major Iranian political groups in Twitter. 
This, in turn, could create incentive for political organizations to engage in opinion manipulation, giving rise to inauthentic activities.

\subsection*{Data}\zlabel{section_Methods_subsection_data}

\textbf{Twitter data.}
We collected tweets by keywords at a time in the vicinity of the peak trendiness of our target topics in Farsi Twitter. 
Our data collection machine uses Twitter's Standard Search API, filtered by keyword and language (Farsi). 
All tweets made available through the API
were collected over a period spanning the rise and fall of the trend.
We used PostgreSQL \cite{PostgreSQL} as a relational database to store references for tweet and user objects and SQLite \cite{SQLite} was used to re-index the objects obtained through the API for each individual topic.
Using API endpoints of Twitter for followers and friends, we also collected followers and friends of users in our data, indexed them in a designated SQLite database, and constructed the edge lists.
If an account, X, is private or deactivated before its followers/friends are requested through the API, it appears in friends/followers list of its follower/friend, Y, if Y is public and active.
However, queries to the API for obtaining the friends/followers of X fail.
We combined the data from friends and followers lists to recover some of the missing links due to an account being private or its deactivation in the gap between our collection of the tweets and that of the friends and followers.

\textbf{Systematic bot detection.}
To perform big data analysis on account types with respect to the typology described in the \ztitleref{section_Methods_subsection_typology} subsection, we use Botometer \cite{Davis_botOrNot, Varol_Botometer, Yang_Botometer}, a supervised machine-learning algorithm, to collect the botscores and Complete Automation Probability (CAP) for the users in our Twitter data. 
We use universal CAP (the language-agnostic complete automation probability) from Botometer results for each account as their botscore.
We include only accounts that were still active (not deactivated or suspended) by the time we collected the botscores.

\textbf{Manual bot detection.}
In an attempt to find an effective and practical approach to using the data obtained from Botometer, we used $1163$ manually annotated randomly-sampled accounts: 
$495$ accounts annotated by members of our group and $668$ more labeled by the participants in a workshop, who were trained on our typology.
The annotations were used to verify the validity of results from Botometer for the purpose of our study and to find a correct threshold for distinguishing between genuine users and likely opinion manipulators. 
Realizing the subjectivity in bot detection, we distributed the accounts among the participants in our workshop in a way that each account is annotated by up to three different participants.
The Kendall's $\tau$ for inter-annotator agreement \cite{kendall1962rank} was approximately $0.67$.
We then chose the CAP intervals such that the variation of labels within each interval is minimized.

\subsection*{User Typology}\zlabel{section_Methods_subsection_typology}

Opinion manipulation in Twitter through different types of bogus accounts is a well-studied subject.
Gorwa et al. comprehensively review prior investigations and typologies for major categories of online bots \cite{Gorwa_typologyGuideToSocialBots}.
They group bogus online accounts into six main types: Web robots, chatbots, spambots, social bots, sockpuppets and trolls, and cyborgs and hybrid accounts. 
Our study concerns accounts that fall in the latter half of these categories - social bots, sockpuppets, and cyborgs.
Comparing the findings of Chu et al. \cite{Chu_humanBotCyborg} with that of Subrahmanian et al. \cite{Subrahmanian_DARPAbotChallenge} shows the significant improvement of Twitter bots in imitating human behavior, a fact confirmed in previous studies \cite{Cresci_paradigmShiftOfBots, Luceri_evolutionOfTwitterBots, Mittal_automatedHighImpact, Subrahmanian_DARPAbotChallenge}, as well as our manual observations.
Therefore, inauthentic accounts could be extremely difficult to detect by a simple set of measures such as temporal activity patterns, and a more complex combination of measures are necessary for distinguishing them from genuine human users.
In this subsection, we introduce a taxonomy of users, with respect to their level of automation, authenticity, and opinion manipulation behavior, which we use to analyze each discussion space. 
Based on our observations through manual exploration of the Farsi Twittersphere, we group the users into three main groups: 
\begin{enumerate}[label=\textbf{\Alph*.}]
    \item \label{a} \textbf{Automated and semi-automated bots:} 
    Twitter bots, often part of a bot squad \cite{Blum_botsInFlocks, Echeverria_botnet}, are used to perform structurally repetitive tasks at a noticeably higher rate compared to humans.
    These are Twitter accounts that could be fully automated without direct involvement of a human, or they can be semi-automated.
    
    \item \label{b} \textbf{Bot-assisted humans or human controlled \textit{campaigning} accounts:}
    These accounts, although may seem like an account belonging to an ordinary person, show a strong similarity to a campaigning account. 
    The content generated by these accounts is strongly oriented towards supporting the standpoint of a group, as one expects to see from an account which belongs to a campaign. 
    These accounts can be divided into the following subcategories.
    \begin{enumerate}[label*=\textbf{\arabic*.}]
        \item \label{b1} \textbf{Deployed users:} 
        These are agents that are deployed to make and control accounts and increase the representation of the stance of an organization.
        Such accounts are often referred to as trolls \cite{Tucker_reviewSocialMediaBots}. 
        Their behavior is typically hardly distinguishable from that of an official campaign or a propaganda news agency, apart from the user information.
        They could be bot-assisted, i.e. equipped with a machine to enhance their performance. 
        The main characteristics of these accounts that separates them from Automated and semi-automated bots (group \ref{a}) and Unusually dedicated users (group \ref{b2}) are that, unlike Automated and semi-automated bots, most often they show a cognitive ability deemed to be exclusively possessed by humans, yet unlike Unusually dedicated users, personal content is noticeably absent in their activities.
        \item \label{b2} \textbf{Unusually dedicated users:} 
        This group consists of users whose accounts seem dedicated to supporting a cause or an organization. 
        They differ from deployed users in that they show personal activities in their tweets significantly more often, e.g. occasionally tweet personal content, engage in personal interactions with others, or express beliefs that do not fit within the main-stream agenda of the organization they side with.
        Their general behavior is nevertheless similar to that of Deployed users (group \ref{b1}).
    \end{enumerate} 
    \item \label{c} \textbf{Genuine users:} The users in this group are humans using Twitter to engage in social activities and express their stance. 
    Apart from showing human-like cognitive behavior, the users belonging to this group often have less homogeneous activities.
    Multidimensional characteristics, personal content, and heterogeneous support of others' stance in a discussion space are among the most outstanding characterizing features of the accounts in this category.
\end{enumerate}
{
Notice that users that fall in the Automated and semi-automated bots category, as well as those we refer to as Deployed users, could be hybrid accounts controlled partially by bots and partially by humans. 
However, the hybrid accounts that are Automated and semi-automated bots are human-assisted bots, while Deployed users are bot-assisted humans, i.e. the former are accounts whose performance in cognitive and content-related aspects is improved by humans and the latter are accounts utilizing machines to enhance the quantitative aspects of their activities.  

Given this typology, we expect the reflection of authentic users on Twitter space to be mostly contained in group \ref{c} —Genuine users— and be limited to Genuine users and Unusually dedicated users.
Unusually dedicated users are where we expect most of the error in bot-detection to lie. 
Many accounts belonging to this group may be indistinguishable from those in deployed or genuine users. 
Neither automated nor manual examination of the accounts could detect structural or content-based features that are exclusive to this group. 
Therefore, it is impractical to label any account as one belonging to Unusually dedicated users with a high degree of confidence. 
}

In our analysis, we target automated and semi-automated bots (group A) and compare their activities across the discussions we analyze against that of genuine users (group C).
In particular, in our analysis of inauthentic activities in Farsi Twitter, group A serves as the target group, group C as the observation group, and group B is primarily a buffer zone which provides a safety margin separating the inauthentic automated agents from authentic accounts.
Although in principle one could study group B as a target group itself and such a study could have useful implications, that does not fit within the objectives of this paper, and we do not conduct a substantial analysis of activities of users in this group.

\begin{figure}[t]
	\centering
	\captionsetup{width=.9\linewidth, format=hang}
	\includegraphics[width=.5\linewidth]{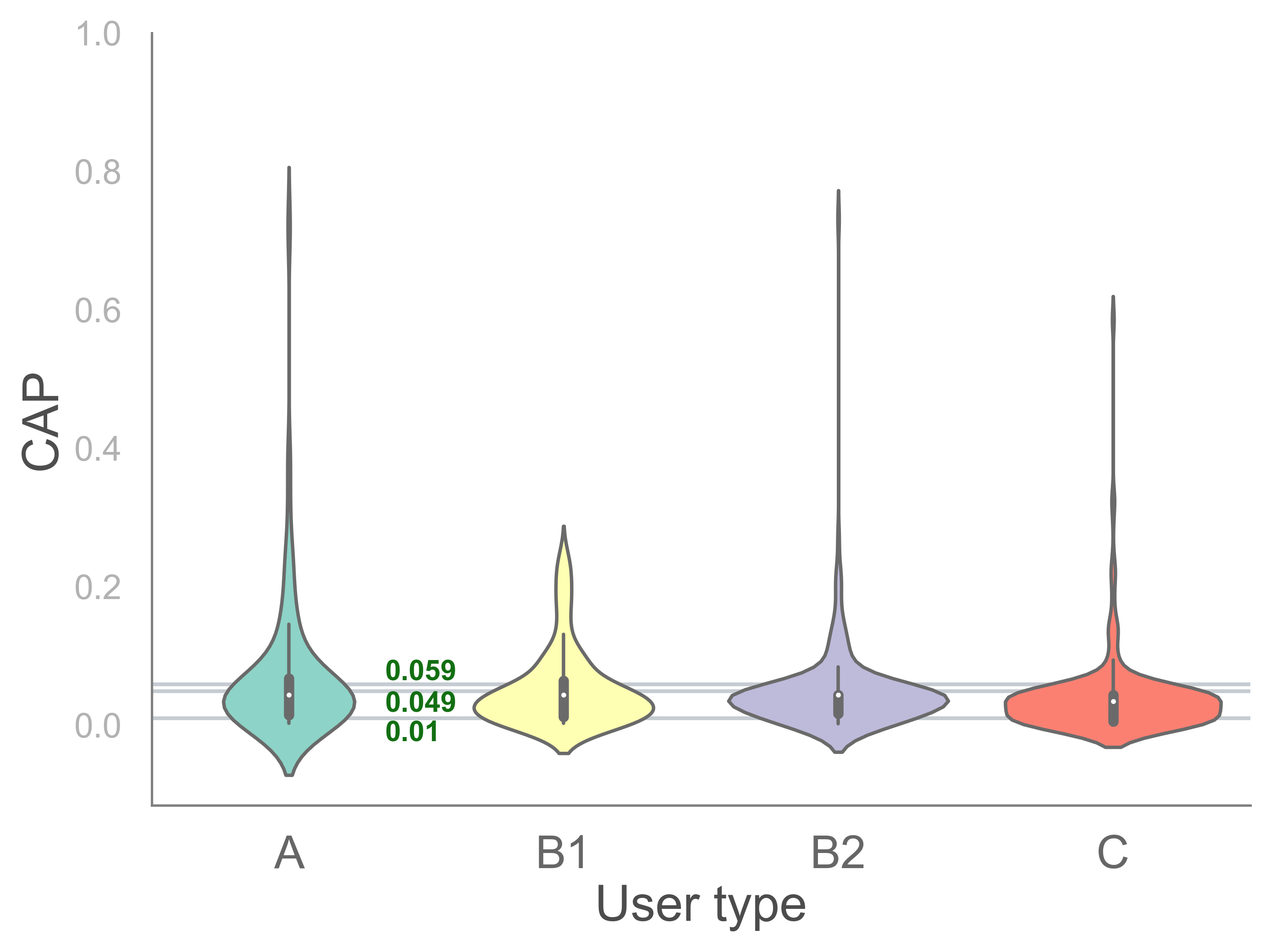}
	\caption{ \footnotesize The distribution of CAP in each group of 668 manually labeled users. 
	The separation lines are at CAP values 0.010, 0.049, and 0.059.}
	\captionsetup{labelformat=empty}
	\label{fig:crwowsroucing_CAPs}
\end{figure}

Using bot-detection methods systematically on big data for mapping users to one of these categories is a challenging task, given the error in bot detection.
As the Botometer team warn users of their bot detection tool, Botometer is rather a complement for human judgement and cannot be relied on for classifying a Twitter account as bot or human \cite{Botometer}.
We, however, verify that, in average, Botometer provides a working estimate that can be used to group users by the types defined in this section. 
Figure \ref{fig:crwowsroucing_CAPs} shows the distribution of CAP in each of the three groups, with group \ref{b} divided into its two subgroups, for manually annotated accounts.
A wide range of CAP is seen for users in all groups, and significant overlap confirms the anticipated misjudgements and errors.
Nevertheless, both mean and median decrease from automated and semi-automated bots to genuine users, which indicates that, at an aggregate level, we can obtain meaningful estimates from Botometer on the behavior of users with respect to the criteria related to opinion manipulation. 
Therefore, although using Botometer for micro-scale analysis of bot-like behavior or labeling of individual accounts according to our typology should be strictly avoided, Botometer can be used to make meaningful implications for macro-scale analysis of opinion manipulation in Twitter when dealing with big data.

\subsection*{Temporal Analysis}\zlabel{section_Methods_subsection_timeSeries}

In order to study the emergence of a trend in Farsi Twitter, we extract time-series data from the tweets that would help us analyze the dynamics of the discussion space. 
This is done through dividing the time interval where the participation in the discussion is significantly higher than before or after that interval into a number of intervals of equal length ($24$ hours) and binning the data into these smaller intervals. 
The data in each time interval is further binned into three user types described in the \ztitleref{section_Methods_subsection_typology} subsection, using their CAP scores. 
We then study the dynamics of the share of each user type in the discussion over time by computing number of tweets tweeted by users from each type within each time interval.
The results, discussed in the \ztitleref{section_results_subsection_participationPatterns} subsection, reveal qualitative differences between apolitical and divisive political discussions with respect to the temporal dynamics of participation in the discussion.

\subsection*{Network Analysis}\zlabel{section_Methods_subsection_networkAnalysis}

We form two types of networks for each discussion: friendship network, and retweet network. 
In both networks the nodes are the users participating in the discussion, i.e. tweeted or retweeted a post containing a corresponding keyword.
In a friendship network there is an edge from user $j$ to user $i$ if $j$ follows $i$, while in a retweet network, a directed edge $(j, i)$ indicates that $j$ has (at least once) retweeted $i$.
Both networks are directed, however, depending on our analysis, we may ignore the direction of the edges. 
For each topic, both types of networks are restricted to the corresponding discussion.
For analyzing communities in our networks, we use the Louvain clustering algorithm \cite{Blondel_Louvain} to detect communities and bin the relevant data by the induced subgraphs.
Note that different runs of the Louvain algorithm could yield different results in a fixed number of iterations.
Considering the use of community structures in our data analysis, we perform a sanity check, computing similarities between several random runs of the Louvain clustering for each network in our dataset, which is included in Appendix A.
Our validation analysis reveals that different runs of the clustering algorithm consistently output highly similar community structures, which confirms the validity of our analysis and interpretations, considering that we only draw aggregate-level conclusions about our observations.

Although retweet networks are more commonly studied for analyzing Twitter discussions, the structure of the friendship network of participants in a discussion could be indicative of various qualitative properties of the discussion space (see e.g. the study by Gonçalves et al. \cite{Goncalves_dunbarsNumber}).
Importantly, the friendship network is the primary medium for circulation of content, and as such,
its structural properties could reveal potential patterns of exposure to content, such as conduciveness to polarization and echo chambers, as we discuss in the \ztitleref{section_results_subsection_content} subsection.
Therefore, analyzing the structure of friendship networks can help us connect the content of discussion to the structure of the corresponding network. 
We discuss an example of how this connection can be studied from a joint analysis of friendship networks and the text of tweets in the \ztitleref{section_results_subsection_content} subsection, where we analyze word clouds and topics corresponding to different friendship communities.
Additionally, since the retweet network is formed through resharing tweets, which is an action users take considering the content of the tweets, the retweet network is inherently content related.
Thus, a comparative analysis of the retweet and friendship networks gives a clearer picture of the relationship between the content and structure of discussion spaces in Twitter, as it is discussed in the \ztitleref{section_results_subsection_echochambers} subsection.

In order to perform the comparison between retweet and friendship networks, we compare the community membership of users with respect to the clustering of each network type.
In a network with segregated communities, the adjacency matrix could be represented as a block-diagonal matrix through a permutation of rows and columns such that nodes belonging to the same community are grouped together.
In any given network, the more segregated the communities are, the closer will such a permutation of the adjacency matrix look to a block-diagonal matrix, with denser diagonal blocks and fewer non-zero off-diagonal entries.
We use this observation in Fig. \ref{fig:RtNw_with_RtComm_vs_FrComm} to visually compare the community structure of the retweet networks with the partitioning induced by friendship communities.
Moreover, we use two clustering similarity measures---Jaccard index and element-center clustering similarity \cite{gates2018element, gates2019clusim}--- to quantify this observation.
To assess the similarities of the retweet and friendship communities, we first form an ensemble of $20$ clusterings for each network, obtained from $20$ runs of the Louvain clustering algorithm using different random seeds.
This yields 400 pairwise similarity values for each discussion space, with respect to each of the two similarity measures, which we then pool by discussion type to compare apolitical and divisive political discussions.
The distributions of the similarity values allow us to draw conclusions on the structural potential of the retweet and friendship networks for yielding similar or dissimilar community structures, as we discuss in the \ztitleref{section_results_subsection_echochambers} subsection.
Performing this analysis over multiple runs of the clustering algorithm allows us to comment on the overlap between retweet and friendship communities despite the randomness in the algorithm and sensitivity of the similarity measures to changes in community memberships.

In order to further quantify the polarized structure of the networks, we compute network modularities \cite{newman2006modularity} with respect to the communities obtained from 100 runs of the Louvain clustering algorithm, to obtain a distribution of modularity values for the networks in each type of discussion.
The modularity of a network with adjacency matrix $A$ is defined as follows
\begin{align*}
    Q &= 
    \frac{1}{2m} \sum_{(v, u) \in V \times V} \left[A_{v, u} - \frac{k_v \ k_u}{2 m} \right] \delta_{c_v, c_u}
\end{align*}
where $m$ is the number of edges in the network, $V$ is the set of nodes, $k_v$ is the degree of node $v$, $c_v$ is the cluster node $v$ belongs to, and $\delta_{c_v, c_u}$ is the Kronecker delta. 
Modularity is a commonly used measure for evaluating the degree of community seggregation and potential for polarization in social networks \cite{Conover_politicalPolarizationTwitter, livne2011party}. 
As we discuss in the \ztitleref{section_results_subsection_echochambers} subsection, our assessment of the distribution of modularity values confirms our findings from analyzing the overlap between retweet and friendship communities.

In addition to the communities, we perform two additional analyses of the meso-scale structure of the networks using core-periphery and bow-tie structures.
Previous studies point to a connection between the core-periphery structure and propogation of content on social networks \cite{saxena2015understanding, barbera2015critical}.
In the core-periphery analysis, the users are partitioned into two groups of core---a densely connected cohesive block of nodes--- and periphery---a sparser and relatively more distant set of nodes \cite{borgatti2000models}.
We use the algorithm proposed by Kojaku and Masuda \cite{kojaku2017finding}, which allows for detecting multiple cores and peripheries within the same network,  in order to find the core-periphery structure of each network in our dataset.
The bow-tie structure on the other hand, yields an alternative partitioning of the nodes in directed networks into seven components \cite{broder2000graph, yang2011bow, jeroenvldj2019}. Recent studies have found connection between the bow-tie structure and discursive communities in online social networks \cite{mattei2022bow}.
We use three components of interest: A strongly connected component (`S'), the nodes at the tail of out-going edges to the S component which act as its `IN' gate, and the nodes at the head of in-coming edges from the S component which act as its `OUT' gate.
In our friendship networks, the `S' component is a densely connected group where there exists a directed path in both directions between every pair of nodes, and as such, it is particularly important for robust propagation of content.
The IN and OUT components on the other hand could be thought of as primary consumers and sources of content in a friendship network, given that an edge in our friendship networks runs in the opposite direction of exposure, from follower to friend.

\subsection*{Content Analysis}\zlabel{section_Methods_subsection_contentAnalysis}

In order to get an overall view of the dominant content, we plot word clouds visualizing frequent words in the tweets. 
To construct the word clouds, we first process the text of the tweets and apply appropriate filters.
We partition the words pool obtained through this process by subpopulations of interest in order to compute word counts for each subpopulation. 
For demonstrating differences between the tweet contents of subpopulations $i$ and $j$, we plot the word cloud corresponding to words that are among the top 5\% most frequent words in $i$ but not among the top 10\% of $j$, and vice versa.
This allows us to observe existing differences between the dominant content of tweets posted by users belonging to different subpopulations.
As discussed in the \ztitleref{section_results_subsection_content} subsection, this analysis is performed once when the grouping of users is done by user types, and once when users are grouped by friendship network communities.

In the interpretation of the results described in the \ztitleref{section_results_subsection_content} subsection, we use the word clouds corresponding to tweets by groups of users as a proxy for the common concern and vocabulary of that group (see \cite{Amor_commWordClouds} for an example of a similar approach).
To further investigate common concerns, we detect salient topics in the set of tweets from each group of users, and consider the most representative words among the top-4 most salient topics in each discussion space.
In order to obtain the topics and their representative words from our tweets, we use BERTopic \cite{grootendorst2022bertopic}, an unsupervised transformer-based topic modeling tool, which has been shown to be well-suited for topic modeling on Twitter posts \cite{egger2022topic}.
Note, as we explain in the \ztitleref{section_results_subsection_content} subsection, when comparing the content of tweets of users in different friendship communities, the network clustering is solely based on the structure of friendship networks, while the text analysis is blind to the network structure. 
Thus, deviation from randomness in the results is an indication of the presence of a relationship between content and structure.
Our results, described in the \ztitleref{section_results_subsection_content} subsection, further suggest that such deviations relate to the topic of the discussion and the position of the participants on the relevant socio-political spectrum.


\section*{Acknowledgements}

In the initial phase of this project, the authors were given access to resources in the Center for Complex Networks and Social Data Science overseen by Dr. Gholamreza Jafari.
We thank Dr. Kosar Karimipour for many fruitful discussions and her constructive suggestions.
We thank Arya Gholampour and Mahdi Sarikhani for organizing and facilitating the workshop where the manual annotation for this project took place.
The authors cordially thank Dr. Gholamreza Jafari, Dr. Kosar Karimipour, Mahdi Sarikhani, and Arya Gholampour for their helpful comments and suggestions.
We sincerely thank the editor and anonymous reviewers for their careful review of the manuscript and their constructive comments, which greatly helped us to improve the quality of this paper.
A major part of this project was done when A.F. was affiliated with and supported by the Department of Engineering Sciences and Applied Mathematics at Northwestern University in Evanston, Illinois and the Max Planck Institute for Mathematics in the Sciences in Leipzig, Germany, and S.M. was affiliated with the Center for Complex Networks and Social Data Science and the Department of Physics at Shahid Beheshti University in Tehran, Iran.
A.F. is currently affiliated with and supported by the Department of Political Science at Duke University in Durham, North Carolina.
S.M. is currently affiliated with the School of Mathematics and Statistics at University College Dublin in Dublin, Ireland, and supported by the Science Foundation of Ireland Centre for Research Training in Foundations of Data Science.

\section*{Author contributions}

A.F. and P.M. designed research and all authors contributed to conceptualization;
P.M. designed and developed the data collection machine;
P.M. and A.F. collected data;
A.F., P.M., Z.P. and S.M. curated data and implemented the computer codes;
A.F. designed the methodology and performed the data analysis;
A.F. and S.M. prepared and visualized data;
All authors contributed to the interpretation of results;
A.F. and A.S. wrote the main body of the manuscript and all authors contributed to its writing and preparation.

\section*{Data availability statement}
Encrypted tweet IDs are available at \url{https://github.com/afarzam/FarsiTwitter_code/tree/main/tweetIDs} and the CAPs for encrypted user IDs are available at \url{https://github.com/afarzam/FarsiTwitter_code/tree/main/CAPs} along with the encrypted keys and Python programs for decryption.
The data collected for this project is subject to Twitter policies regarding data collection for academic research.
Moreover, for privacy concerns, we are unable to make our dataset available in its original format.
Once these considerations are realized, we may be able to provide the private keys for decryption upon request and further consultation with Twitter.
The code for reproducing the main figures and the preceding data preparation is available at \url{https://github.com/afarzam/FarsiTwitter_code}.


\section*{Additional information}

The authors declare no competing interests.


{\footnotesize
	\bibliographystyle{unsrt}  
    \bibliography{citations.bib}  
}

\newpage

\begin{appendices}
	\section{Other Discussion Topics\zlabel{Appendix_others}}

Throughout the Results section in the main body of this paper, we include figures and discuss results comparing divisive political discussions against apolitical discussions.
In the cases where the analysis is performed for each discussion separately, we include the results for on two discussion topics: Snapp (divisive political) and Valentine (apolitical).
These two topics are included as examples in the main manuscript.
In this appendix, we include figures for the results, which extend to other topics within each discussion type.
Additionally, we also include distributions of complete automation probability (CAP) across all types of discussions, visualizations of all networks, as well as an analysis of the similarity of the results from different runs of the Louvain clustering algorithm on each network.


\subsection*{Distribution of Complete Automation Probabilities
\label{appendix_others_subsection_distributions}} 
\begin{figure}[hbt!]
	\centering
	\captionsetup{width=.9\linewidth, format=hang}
	\includegraphics[width=.4\linewidth]{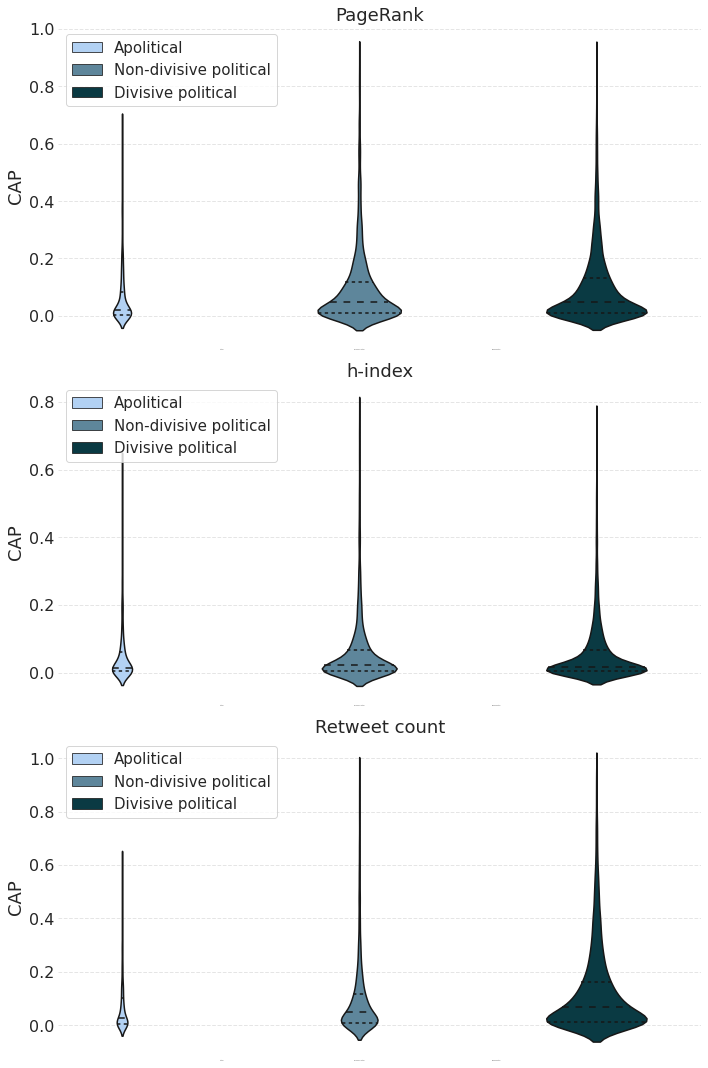}
	\caption{ \footnotesize 
	The distributions of CAP scores among the most influential accounts, by PageRank (top), h-index (middle), and retweet count (bottom), in the retweet network of the Apolitical(left), Non-divisive Political (middle), and Divisive Political (right) discussions.
	The violin plots, and the quartile lines within each violin plot, show the distribution of CAP scores for users among the top decile of accounts with respect to the corresponding measure, in each topic within each category of discussion.}
	\captionsetup{labelformat=empty}
	\label{appendix_fig:appendix_distributions}
\end{figure}

We begin our Results section in the main manuscript with a discussion on prevalence of inauthentic activities across different discussion types.
As we discuss in the manuscript, this paper primarily aims to study inauthentic activities in divisive political discussions compared against apolitical discussions which serve as a benchmark.
The other category of discussion ---non-divisive political discussions--- provide a safety margin, separating the two types of discussions.
To present an overall view of inauthentic activities in this category, in this appendix we include the distribution of Complete Automation Probabilities (CAP) in this type of discussions, for users belonging to the top decile of each of the three influence measures in the retweet network, as we discuss in the manuscript for apolitical and divisive political discussions.
The distributions are shown in Fig. \ref{appendix_fig:appendix_distributions}.

\subsection*{Network Communities and Visualisation
\label{appendix_others_subsection_communityStructure}}

\begin{figure*}[hbt!]
	\centering
	\captionsetup{width=.9\linewidth, format=hang}
	\includegraphics[width=.9\linewidth]{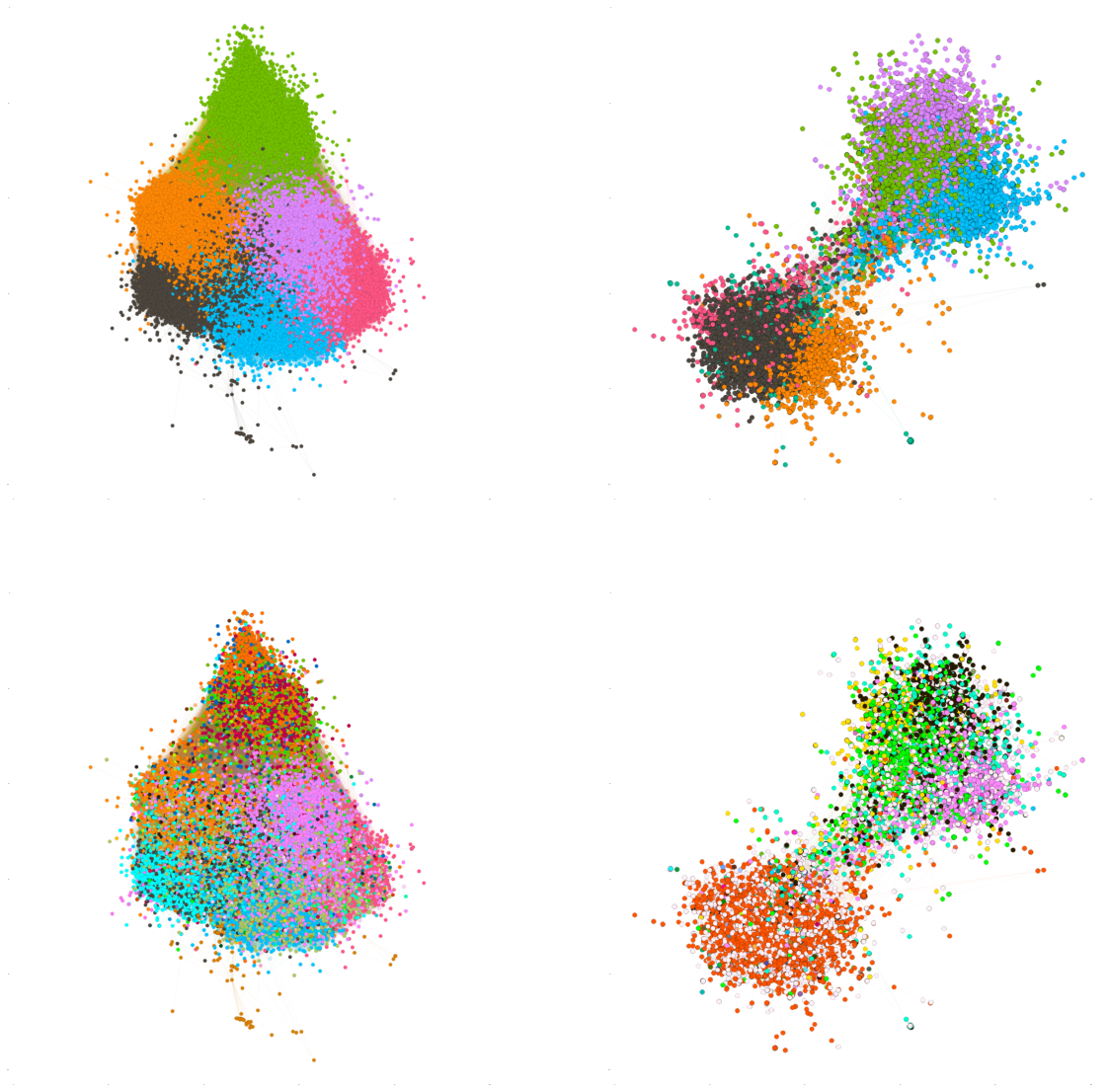}
		\caption{ \footnotesize The retweet and friendship networks for the Twitter discussion about Snapp.
		\textbf{Top left:} The friendship network color coded according the communities in the same friendship network.
		\textbf{Top right:} The retweet network color coded according to the communities in the same retweet network.
		\textbf{Bottom left:} The friendship network color coded according the communities in the corresponding retweet network.
		\textbf{Bottom right:} The retweet network color coded according to the communities in the corresponding friendship network. 
		Excluding the isolated nodes, the retweet network for this discussion has 21029 nodes and 69565 edges, and the friendship network has 22993 nodes and 1728354 edges.
		Small clusters are excluded from this visualization.
		The retweet networks visualized in this figure contain 17483 nodes and 60768 edges, and the friendship networks contain 22991 nodes and 1728352 edges.
		Note that the grey dots, which are more abundant in the bottom row, mark those nodes that were not in any of the major communities.}
		\label{appendix_fig:snapp_network_visualized}
\end{figure*}

\begin{figure*}[hbt!]
	\centering
	\captionsetup{width=.9\linewidth, format=hang}
	\includegraphics[width=.9\linewidth]{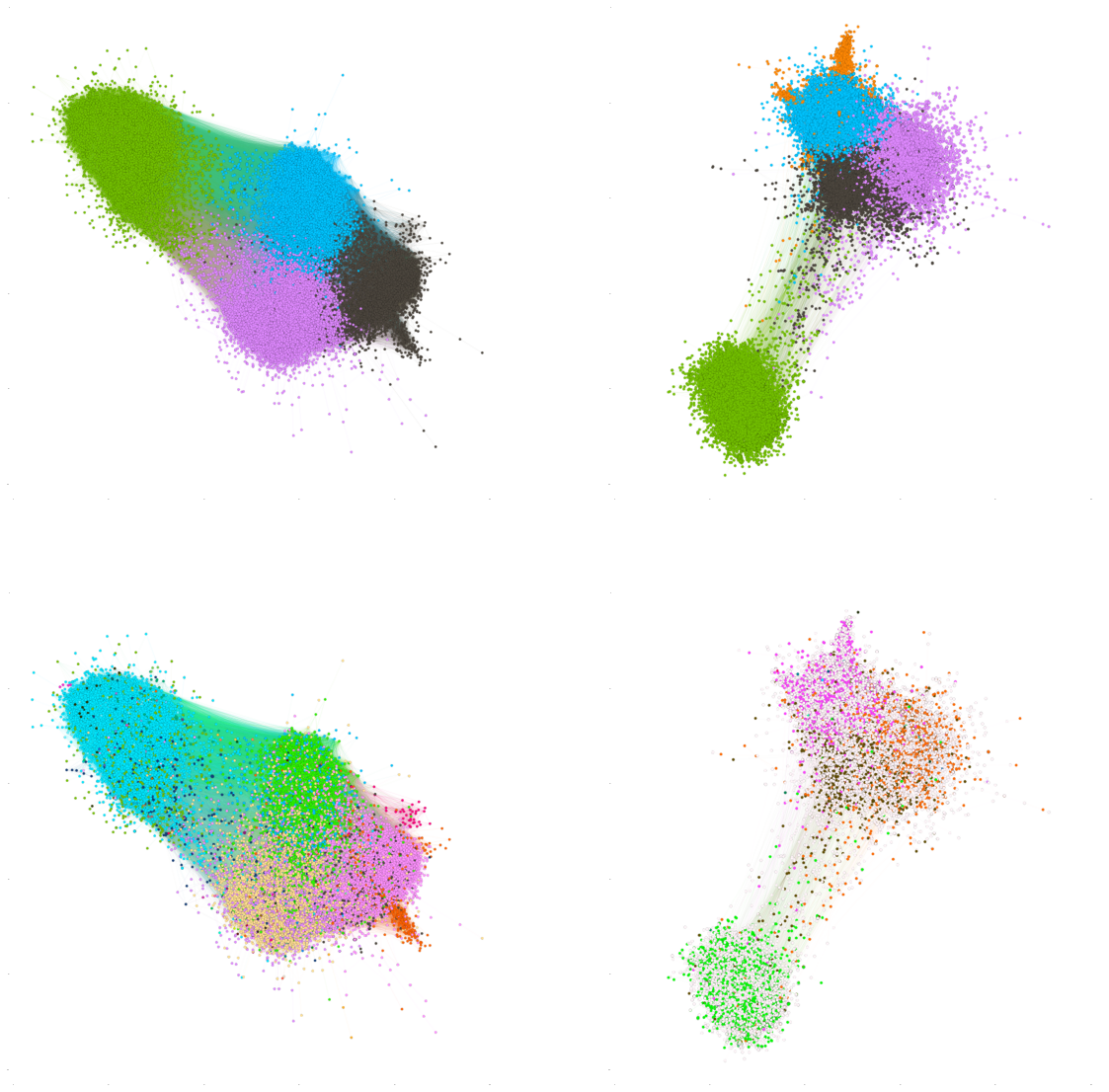}
		\caption{ \footnotesize The retweet and friendship networks for the Twitter discussion about Blue Girl.
		\textbf{Top left:} The friendship network color coded according the communities in the same friendship network.
		\textbf{Top right:} The retweet network color coded according to the communities in the same retweet network.
		\textbf{Bottom left:} The friendship network color coded according the communities in the corresponding retweet network.
		\textbf{Bottom right:} The retweet network color coded according to the communities in the corresponding friendship network. 
		Excluding the isolated nodes, the retweet network for this discussion has 47274 nodes and 320041 edges, and the friendship network has 15898 nodes and 1574559 edges.
		Small clusters are excluded from this visualization.
		The retweet networks visualized in this figure contain 45440 nodes and 315411 edges, and the friendship networks contain 15785 nodes and 1561803 edges.}
		\label{appendix_fig:bluegirl_network_visualized}
\end{figure*}

\begin{figure*}[hbt!]
	\centering
	\captionsetup{width=.9\linewidth, format=hang}
	\includegraphics[width=.9\linewidth]{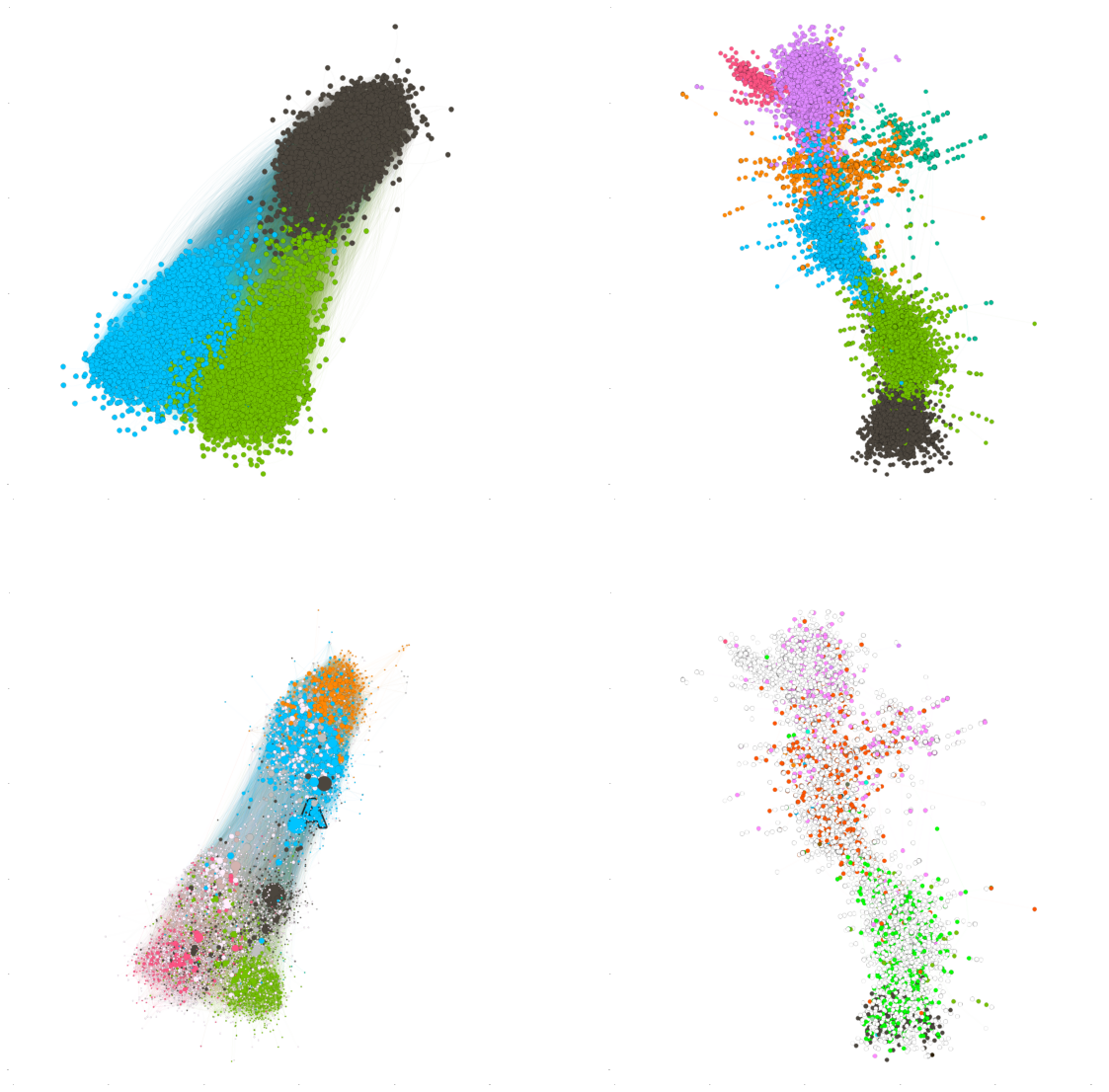}
		\caption{ \footnotesize The retweet and friendship networks for the Twitter discussion about Trump.
		\textbf{Top left:} The friendship network color coded according the communities in the same friendship network.
		\textbf{Top right:} The retweet network color coded according to the communities in the same retweet network.
		\textbf{Bottom left:} The friendship network color coded according the communities in the corresponding retweet network.
		\textbf{Bottom right:} The retweet network color coded according to the communities in the corresponding friendship network. 
		Excluding the isolated nodes, the retweet network for this discussion has 18596 nodes and 44254 edges, and the friendship network has 5372 nodes and 235668 edges.
		Small clusters are excluded from this visualization.
		The retweet networks visualized in this figure contain 15044 nodes and 39667 edges, and the friendship networks contain 5354 nodes and 234544 edges.}
			\label{appendix_fig:trump_network_visualized}
\end{figure*}

\begin{figure*}[hbt!]
	\centering
	\captionsetup{width=.9\linewidth, format=hang}
	\includegraphics[width=.9\linewidth]{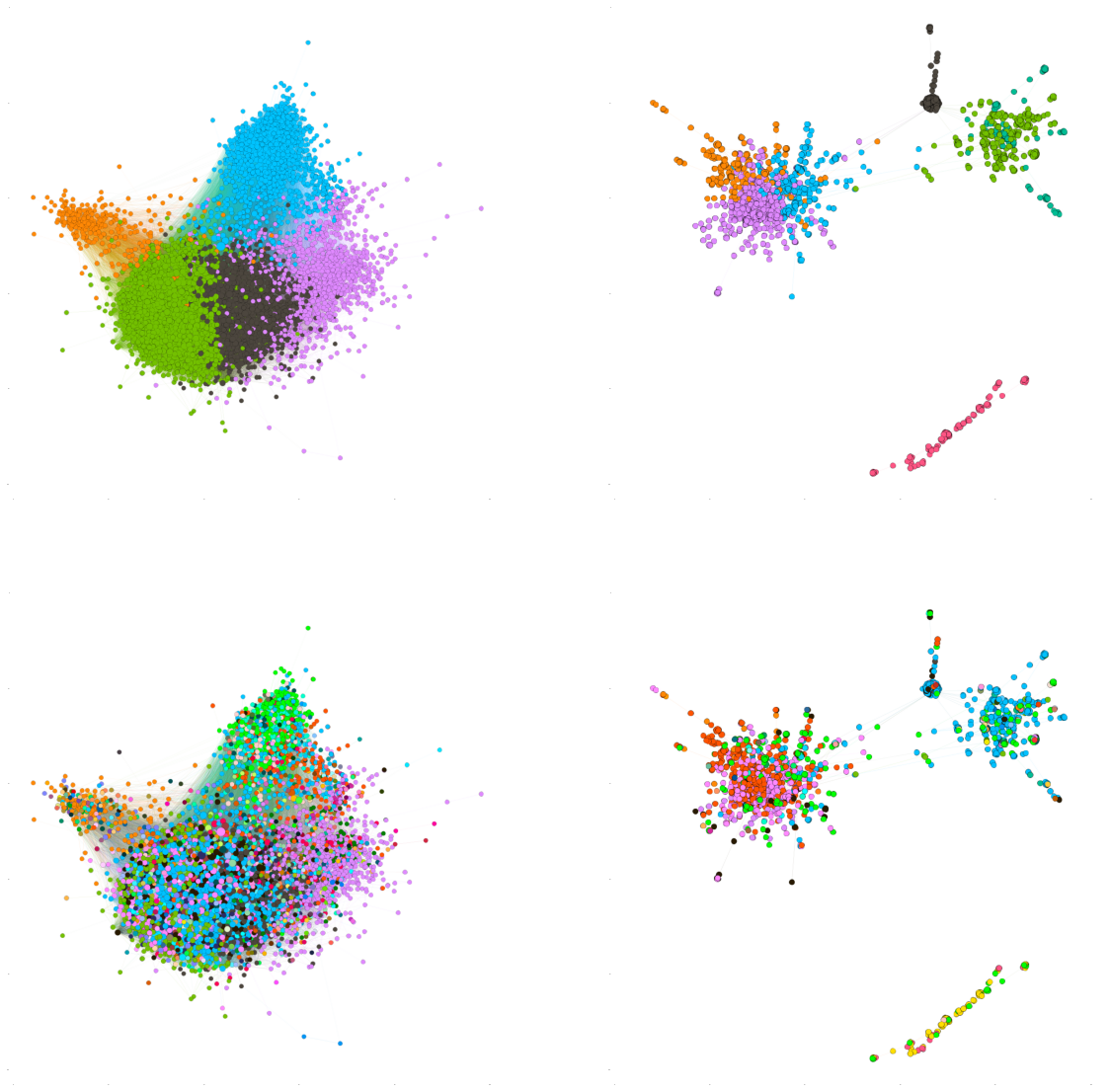}
		\caption{ \footnotesize The retweet and friendship networks for the Twitter discussion about Valentine.
		\textbf{Top left:} The friendship network color coded according the communities in the same friendship network.
		\textbf{Top right:} The retweet network color coded according to the communities in the same retweet network.
		\textbf{Bottom left:} The friendship network color coded according the communities in the corresponding retweet network.
		\textbf{Bottom right:} The retweet network color coded according to the communities in the corresponding friendship network. 
		Excluding the isolated nodes, the retweet network for this discussion has 3705 nodes and 4491 edges, and the friendship network has 4001 nodes and 369965 edges.
		Small clusters are excluded from this visualization.
		The retweet networks visualized in this figure contain 1760 nodes and 2623 edges, and the friendship networks contain 3981 nodes and 369025 edges.}
		\label{appendix_fig:valentine_network_visualized}
\end{figure*}

\begin{figure*}[hbt!]
	\centering
	\captionsetup{width=.9\linewidth, format=hang}
	\includegraphics[width=.9\linewidth]{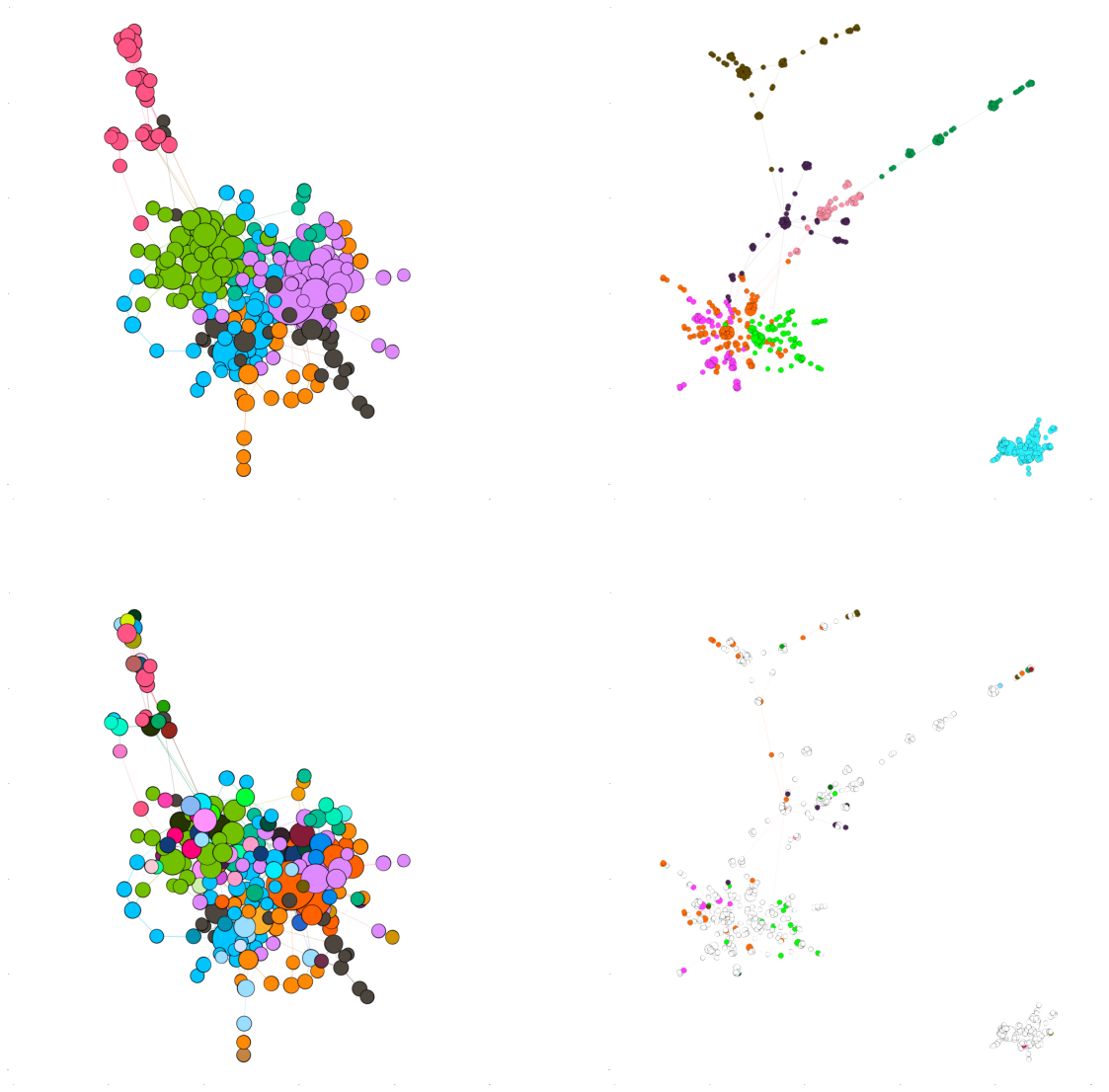}
		\caption{ \footnotesize The retweet and friendship networks for the Twitter discussion about Kobe Bryant.
		\textbf{Top left:} The friendship network color coded according the communities in the same friendship network.
		\textbf{Top right:} The retweet network color coded according to the communities in the same retweet network.
		\textbf{Bottom left:} The friendship network color coded according the communities in the corresponding retweet network.
		\textbf{Bottom right:} The retweet network color coded according to the communities in the corresponding friendship network. 
		Excluding the isolated nodes, the retweet network for this discussion has 1783 nodes and 1748 edges, and the friendship network has 227 nodes and 749 edges.
		Small clusters are excluded from this visualization.
		The retweet networks visualized in this figure contain 1783 nodes and 1748 edges, and the friendship networks contain 211 nodes and 709 edges.}
			\label{appendix_fig:kobe_network_visualized}
\end{figure*}

\begin{figure*}[hbt!]
	\centering
	\captionsetup{width=.9\linewidth, format=hang}
	\includegraphics[width=.9\linewidth]{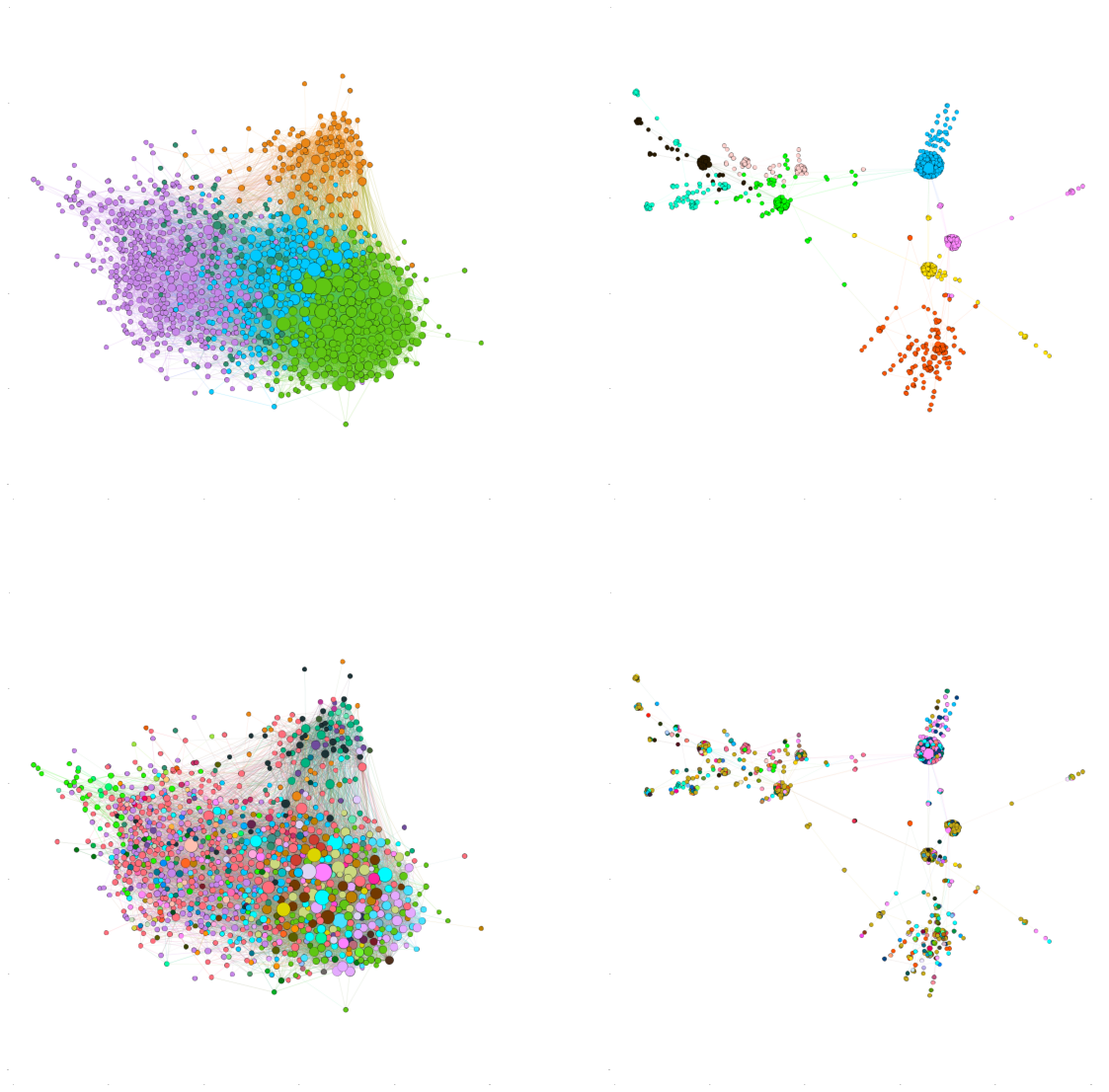}
		\caption{ \footnotesize The retweet and friendship networks for the Twitter discussion about Depression.
		\textbf{Top left:} The friendship network color coded according the communities in the same friendship network.
		\textbf{Top right:} The retweet network color coded according to the communities in the same retweet network.
		\textbf{Bottom left:} The friendship network color coded according the communities in the corresponding retweet network.
		\textbf{Bottom right:} The retweet network color coded according to the communities in the corresponding friendship network. 
		Excluding the isolated nodes, the retweet network for this discussion has 2655 nodes and 2727 edges, and the friendship network has 1113 nodes and 25589 edges.
		Small clusters are excluded from this visualization.
		The retweet networks visualized in this figure contain 1115 nodes and 1230 edges, and the friendship networks contain 1113 nodes and 25589 edges.}
		\label{appendix_fig:afosrdegi_network_visualized}
\end{figure*}

\begin{figure*}[hbt!]
	\centering
	\captionsetup{width=.9\linewidth, format=hang}
	\includegraphics[width=.9\linewidth]{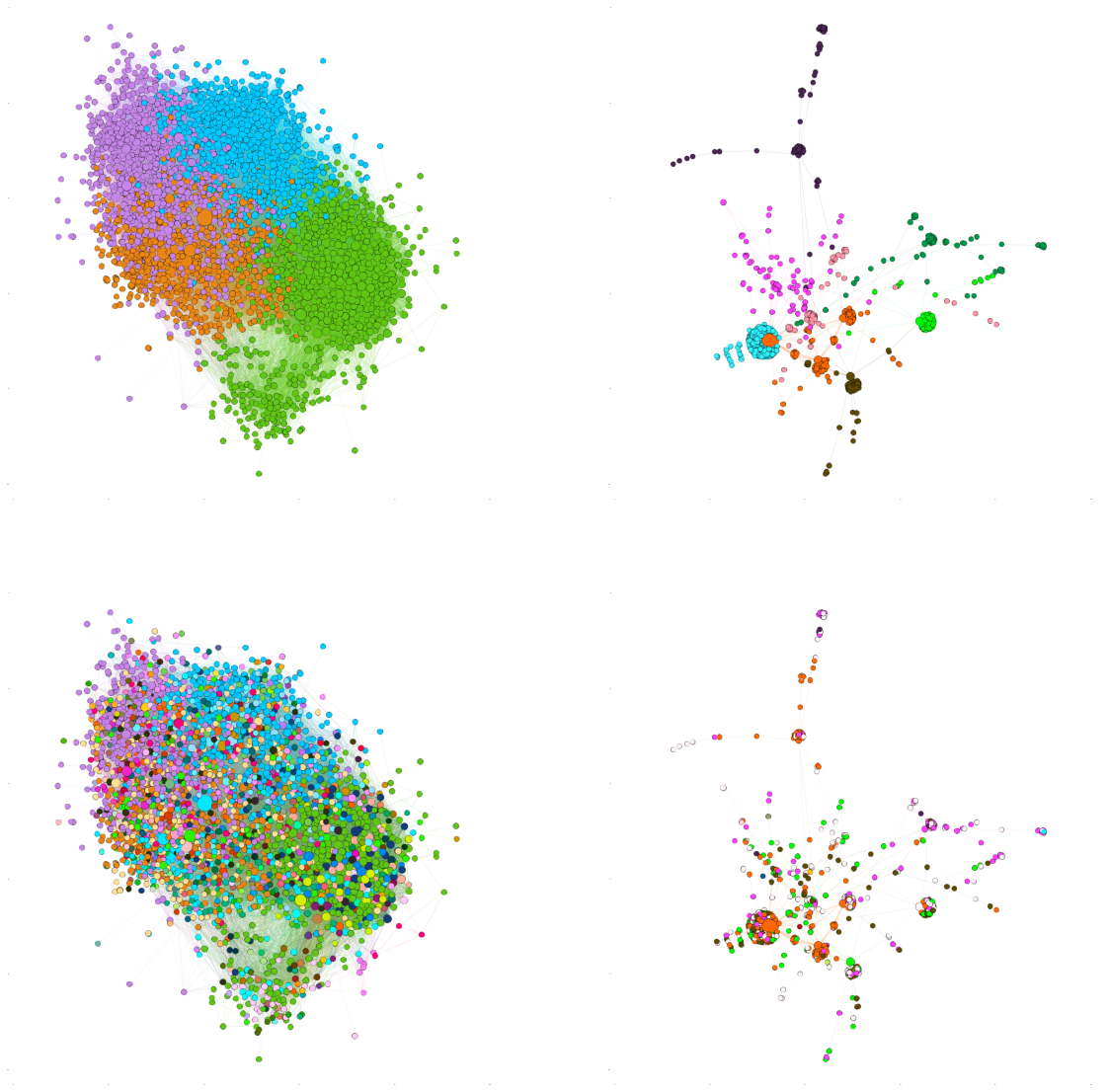}
		\caption{ \footnotesize The retweet and friendship networks for the Twitter discussion about Golab Adineh.
		\textbf{Top left:} The friendship network color coded according the communities in the same friendship network.
		\textbf{Top right:} The retweet network color coded according to the communities in the same retweet network.
		\textbf{Bottom left:} The friendship network color coded according the communities in the corresponding retweet network.
		\textbf{Bottom right:} The retweet network color coded according to the communities in the corresponding friendship network. 
		Excluding the isolated nodes, the retweet network for this discussion has 2118 nodes and 2192 edges, and the friendship network has 2676 nodes and 78339 edges.
		Small clusters are excluded from this visualization.
		The retweet networks visualized in this figure contain 910 nodes and 998 edges, and the friendship networks contain 2668 nodes and 78128 edges.}
			\label{appendix_fig:golab_adine_network_visualized}
\end{figure*}

In this section of the appendix, we include network visualizations for discussions in both target categories, with the nodes color coded by network community, in Figures \ref{appendix_fig:snapp_network_visualized} through \ref{appendix_fig:golab_adine_network_visualized}.
These visualizations show friendship and retweet networks, and for each, we include two visualizations with two different color codings: one reflecting the communities in the same network, and one reflecting the communities in the other network.
That is, for each discussion, we include four visualizations: friendship network with nodes color coded by friendship communities, friendship network with nodes color coded by retweet communities, retweet network with nodes color coded by retweet communities, retweet network with nodes color coded by friendship communities.
Please note that in these visualisations only 7 major communities of networks are demonstrated and smaller, isolated communities are removed to better represent the structure of the main network.
Please further note that accounts that participate in the discussion, but neither retweeted a tweet nor were retweeted by other accounts, correspond to nodes in the friendship network that do not exist in the retweet network.
These nodes are colored white in the visualizations of friendship networks reflecting retweet communities.

\begin{figure}[hbt!]
        \centering
        \captionsetup{width=.9\linewidth, format=hang, labelformat=empty}
          \includegraphics[width=.8\linewidth]{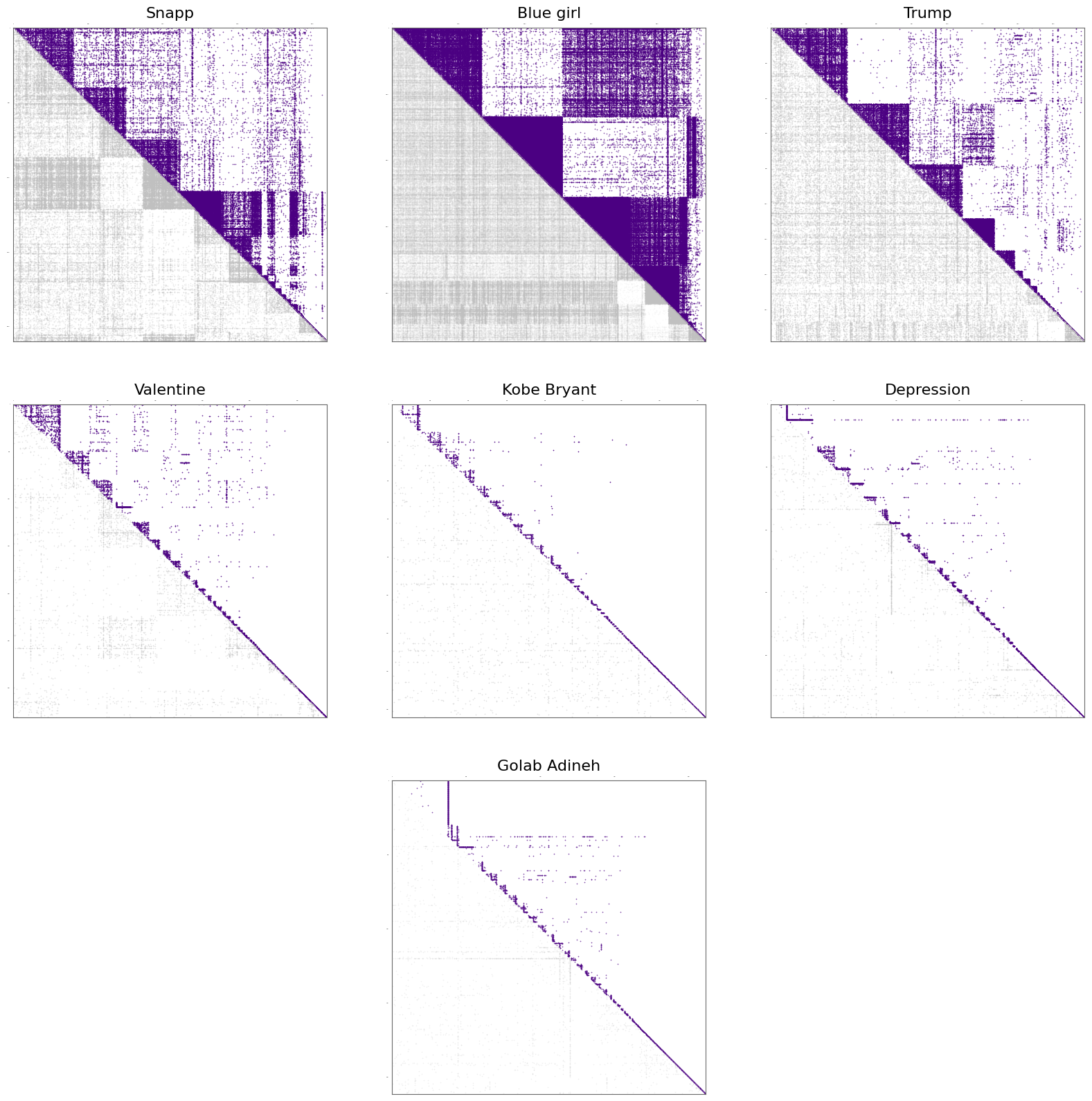}
	\caption{ \footnotesize The symmetrized retweet networks for each of the divisive political (top row) apolitical (the rest) Twitter discussions.
	In the upper triangle the rows and columns are reordered such to group nodes by their retweet communities.
	In the lower triangle, the reordering groups nodes by the community they belong to in the corresponding friendship network.
	Excluding the isolated nodes, the retweet network of the Snapp discussion has 21029 nodes and 69565 edges, and the retweet network of the Valentine discussion has 3705 nodes and 4491 edges.}
	\captionsetup{labelformat=empty}
	\label{appendix_fig:RtNw_with_RtComm_vs_FrComm}
\end{figure}

In the Results section we discuss the distribution of suspended and deactivated accounts in different retweet network communities for divisive political discussions compared against apolitical discussions, and also compared the retweet and friendship network communities.
In particular, we discuss the similarities between retweet and friendship communities in the Inauthentic Activity in Echo Chambers subsection of the main manuscript, where we include visualizations of the network adjacency matrices for the Snapp (divisive political) and Valentine (apolitical) discussions with rows and columns permuted such that nodes belonging to the same community are grouped together.
These sparsity patterns of the adjacency matrices are demonstrated in Fig. \ref{appendix_fig:RtNw_with_RtComm_vs_FrComm} in this appendix.

\begin{figure}[hbt!]
	\centering
	\captionsetup{width=.9\linewidth, format=hang}
	\includegraphics[width=.75\linewidth]{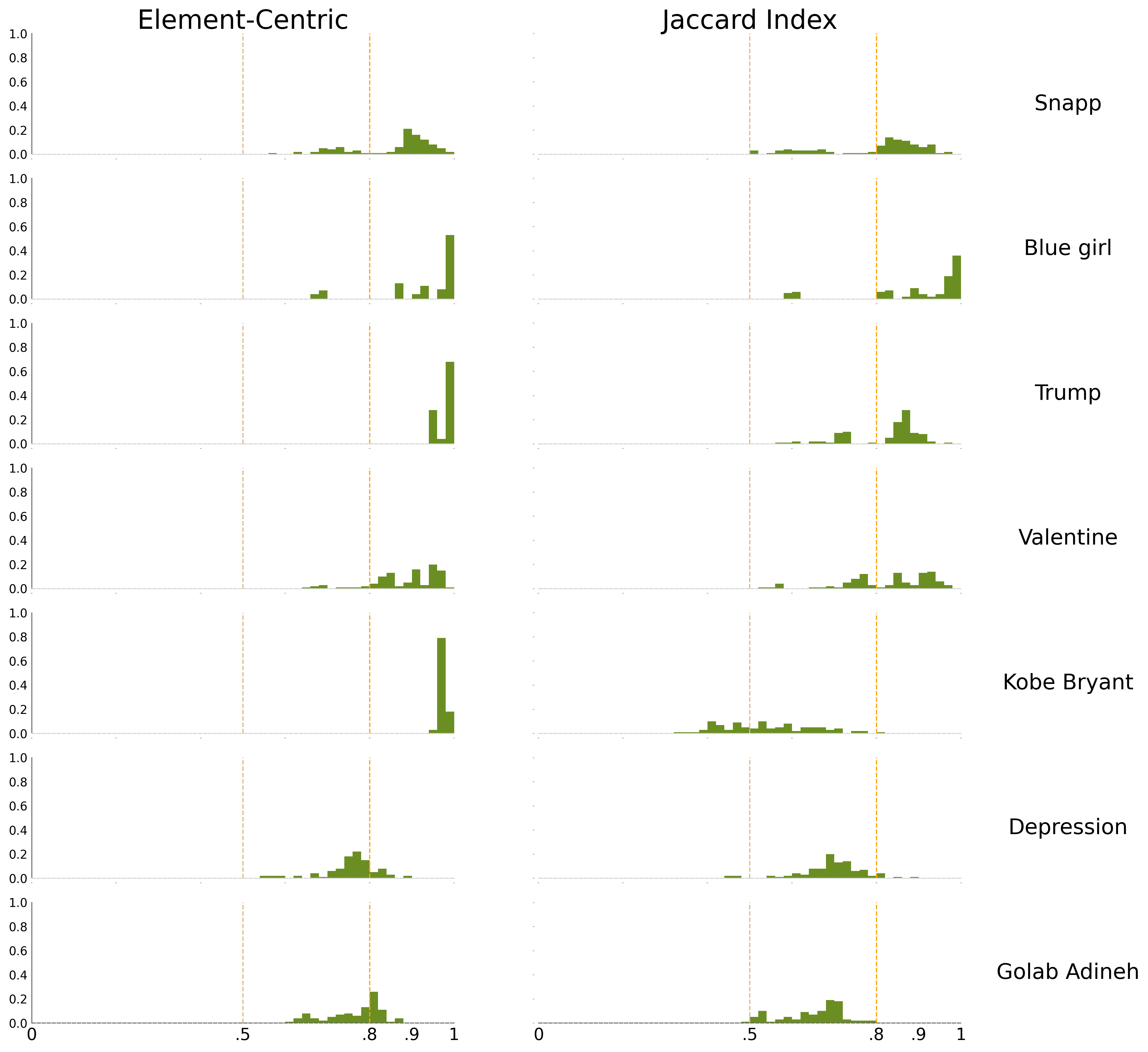}
	\caption{ \footnotesize The distribution of similarity values between the Louvain clustering used in our analysis for the friendship networks and 100 alternative clusterings of the same networks obtained from 100 runs of the algorithms using different random seeds.  
	On the left, the similarities are computed according to element-centric similarity.
	On the right, the similarities are computed according to Jaccard index.}
	\captionsetup{labelformat=empty}
	\label{appendix_fig:clusims_frnd}
\end{figure}

\begin{figure}[hbt!]
	\centering
	\captionsetup{width=.9\linewidth, format=hang}
	\includegraphics[width=.75\linewidth]{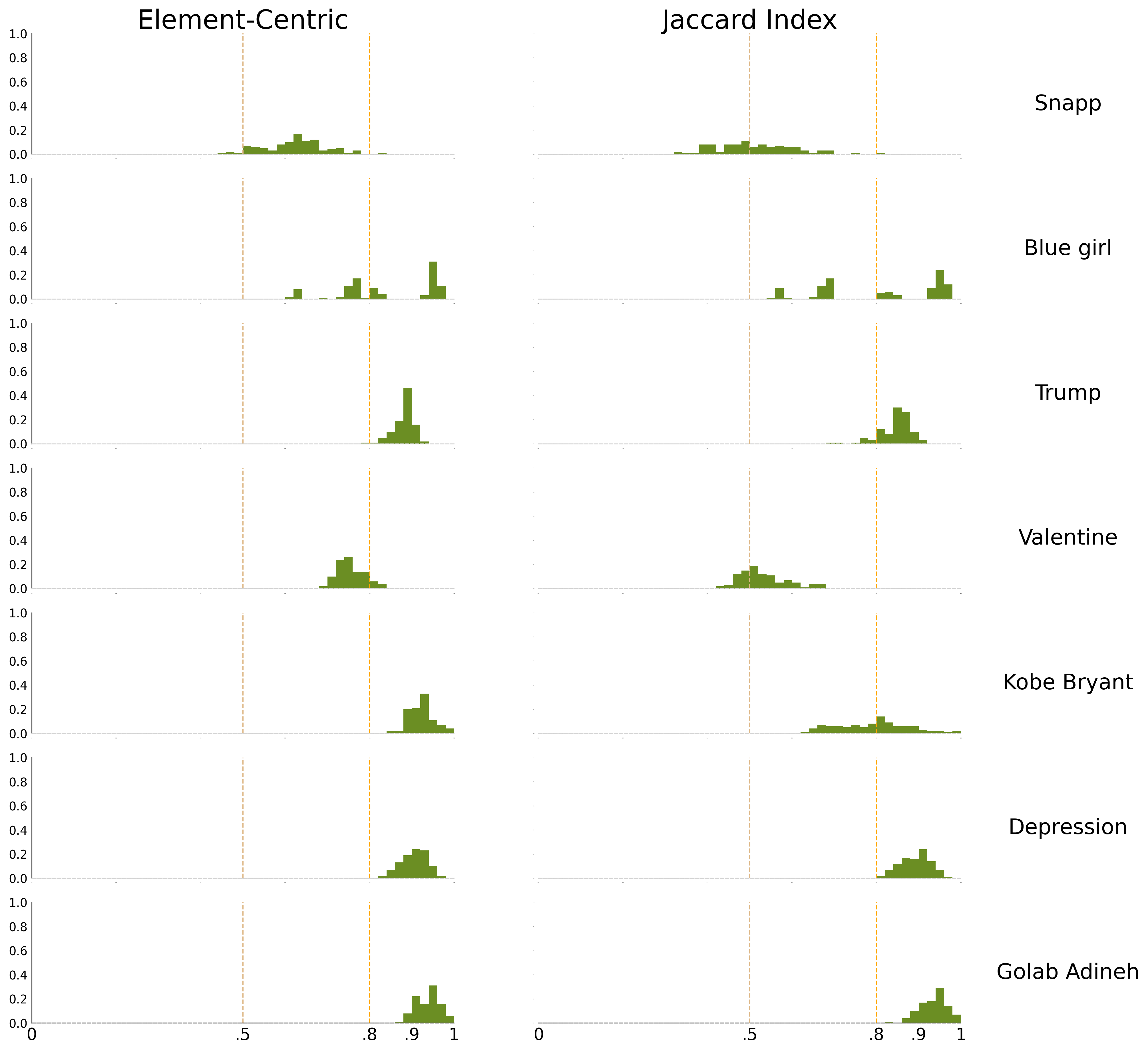}
	\caption{ \footnotesize The distribution of similarity values between the Louvain clustering used in our analysis for the retweet networks and 100 alternative clusterings of the same networks obtained from 100 runs of the algorithms using different random seeds.  
	On the left, the similarities are computed according to element-centric similarity.
	On the right, the similarities are computed according to Jaccard index.}
	\captionsetup{labelformat=empty}
	\label{appendix_fig:clusims_rt}
\end{figure}

As we discuss in the Results as well as the Methods sections of the main manuscript, we use the Louvain clustering algorithm to detect network communities in our data. 
In addition to considering the adjacency matrix structure shown in Fig. \ref{appendix_fig:RtNw_with_RtComm_vs_FrComm}, to further confirm the reliability of our community detection for our macro-scale analysis, we construct an ensemble of 100 clusterings obtained from different runs of the Louvain algorithm.
We then compute the pairwise similarities between the network clustering we use in other analyses and each of the alternative clusterings, with respect to two similarity measures ---Jaccard index and element-centric similarity---, as we explain in the Methods section of the main manuscript.
The results, included in Figures \ref{appendix_fig:clusims_frnd} and \ref{appendix_fig:clusims_rt}, indicate that the vast majority of similarity values fall above $0.8$ (out of a maximum similarity of $1$) with nearly all of the values above $0.5$, which confirms that our observations and interpretations are not sensitive to differences between different runs of the algorithm, considering that our analyses look at the network at the macro scale and draw aggregate-level conclusions.

\subsection*{Participation in Trending Discussions
\label{appendix_others_subsection_participationPatterns}}

\begin{figure*}[hbt!]
	\centering
	\captionsetup{width=.9\linewidth, format=hang}
	\includegraphics[width=.9\linewidth]{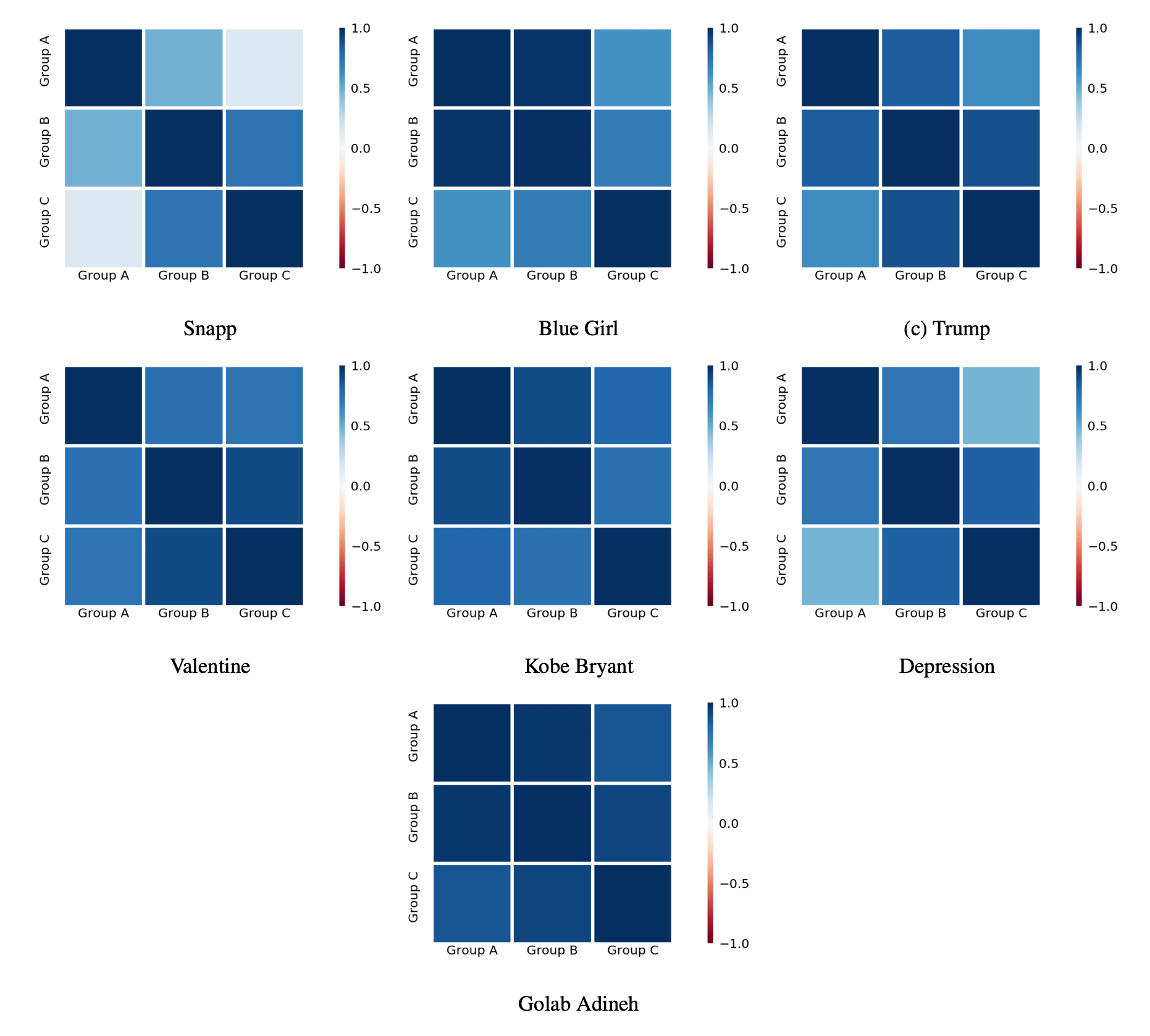}
	\caption{ \footnotesize The Spearman correlation between the changes in the number of tweets from one day to the next for different groups of users — automated inauthentic accounts (group A), bot assisted humans and trolls (group B), and genuine users (group C).
	Please refer to Appendix section \ztitleref{Appendix_topics} for descriptions of discussion topics and the number of tweets for each topic.
	}
	\label{appendix_fig:participationTrend_correlation_t_others}
\end{figure*}

{
We show the differences between the Snapp (divisive political) and Valentine (apolitical) discussions with respect to the temporal pattern of participation by different groups of users in the Results section of the main manuscript.
As it is shown by Spearman correlations visualized in Fig. \ref{appendix_fig:participationTrend_correlation_t_others}, the discussion in the main manuscript generally holds across other discussions of the same types with the exception of the apolitical discussion on Depression, which deviates from this general pattern.
The correlation difference between groups A and C in the Trump discussion is also not as pronounced as it is in the Blue Girl and Snapp discussions, but this is in line with our general understanding about the more mixed collection of tweets on the Trump discussion, as opposed to the more unequivocally divisive discussions about Blue Girl and Snapp.
}

\subsection*{Differences in Content
\label{appendix_others_subsection_content}}

\begin{figure}[hbt!]
	\centering
	\captionsetup{width=.9\linewidth, format=hang}
	\includegraphics[width=.7\linewidth]{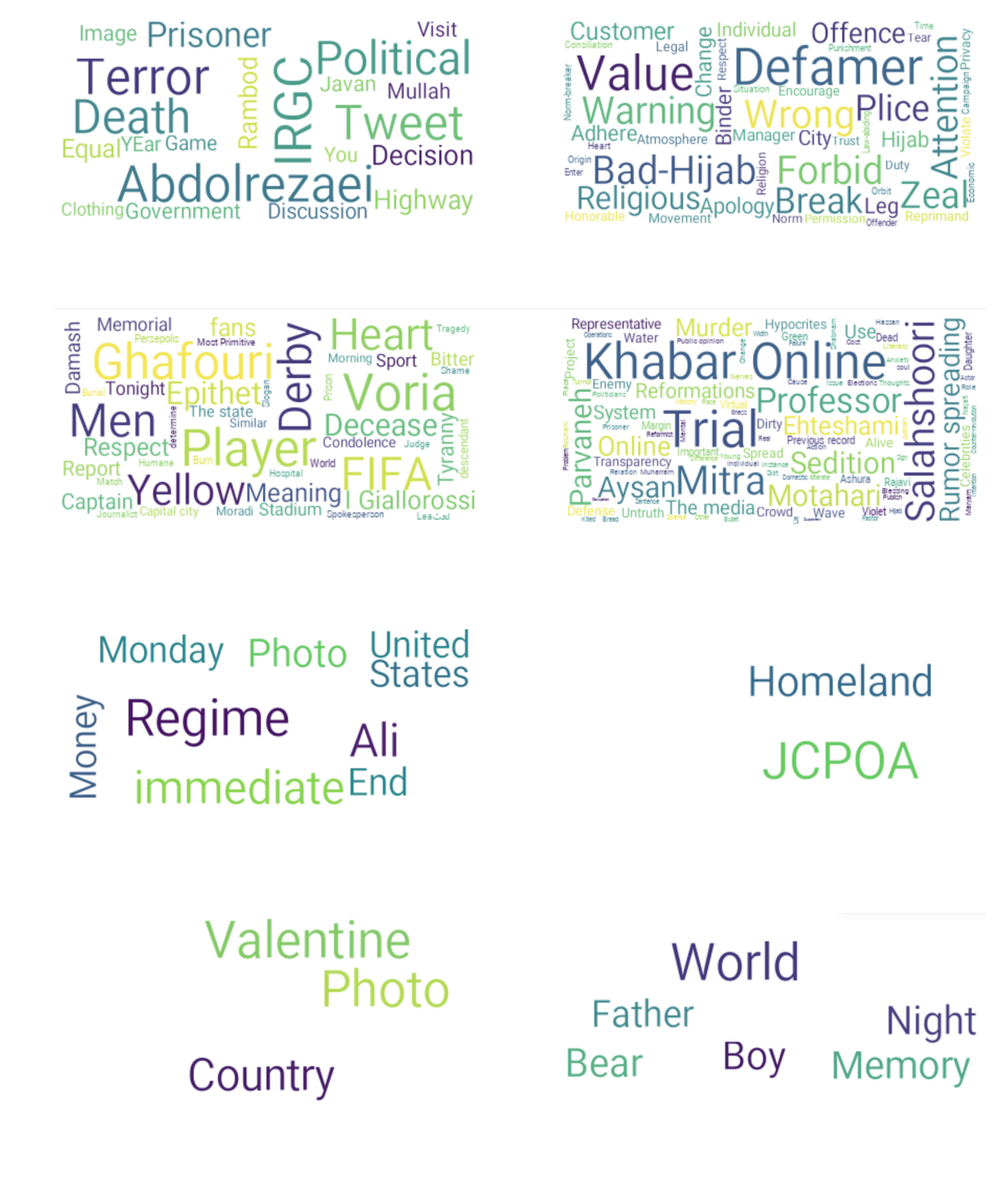}
	\caption{ \footnotesize { Difference word clouds by CAP for the Snapp, Blue Girl, Trump, and Valentine  discussions, from top to bottom.
	Left: top 5\% most frequent words used by users in major community 1 which are not among the top 10\% most frequent words used by users in major community 2. 
	Right: top 5\% most frequent words used by users in major community 2 which are not among the top 10\% most frequent words used by users in major community 1.
	Note that the difference word clouds for the apolitical discussions about Kobe Bryant, Depression, and Golab Adineh were empty, i.e. there is no difference between the two communities, in line with our main results.}}
	\label{appendix_fig:wordclouds_1vs2}
\end{figure}

\begin{figure}[hbt!]
	\centering
	\captionsetup{width=.9\linewidth, format=hang}
	\includegraphics[width=.55\linewidth]{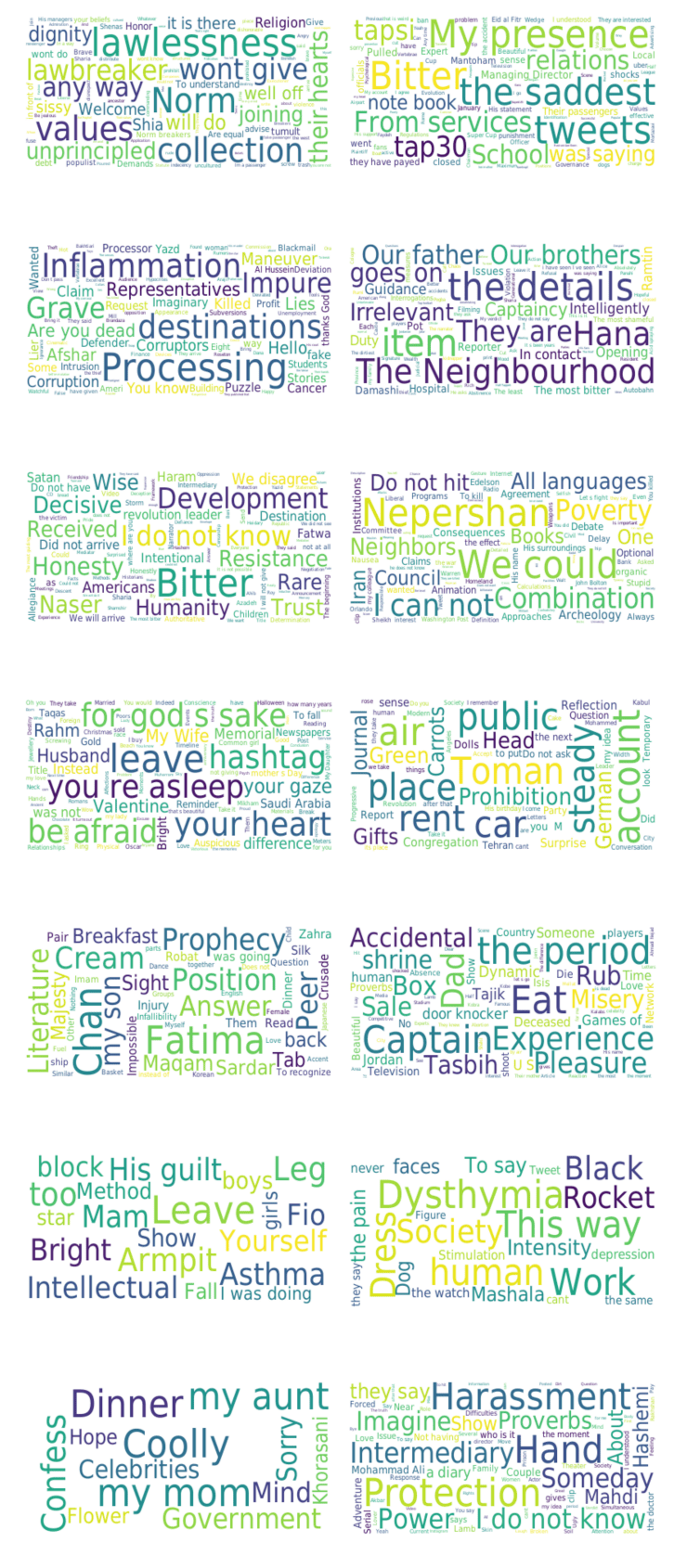}
	\caption{ \footnotesize { Difference word clouds by CAP for the Snapp, Blue Girl, Trump, Valentine, Kobe Bryant, Depression, and Golab Adineh discussions, from top to bottom.
	Left: top 5\% most frequent words across the high-CAP group which are absent from the top 10\% most frequent words used by the low-CAP group. 
	Right: top 5\% most frequent words across the low-CAP group which are absent from the top 10\% most frequent words used by the high-CAP group.}}
	\label{appendix_fig:wordclouds_AvsC}
\end{figure}

\begin{figure}[hbt!]
	\centering
	\captionsetup{width=.9\linewidth, format=hang}
	\includegraphics[width=.85\linewidth]{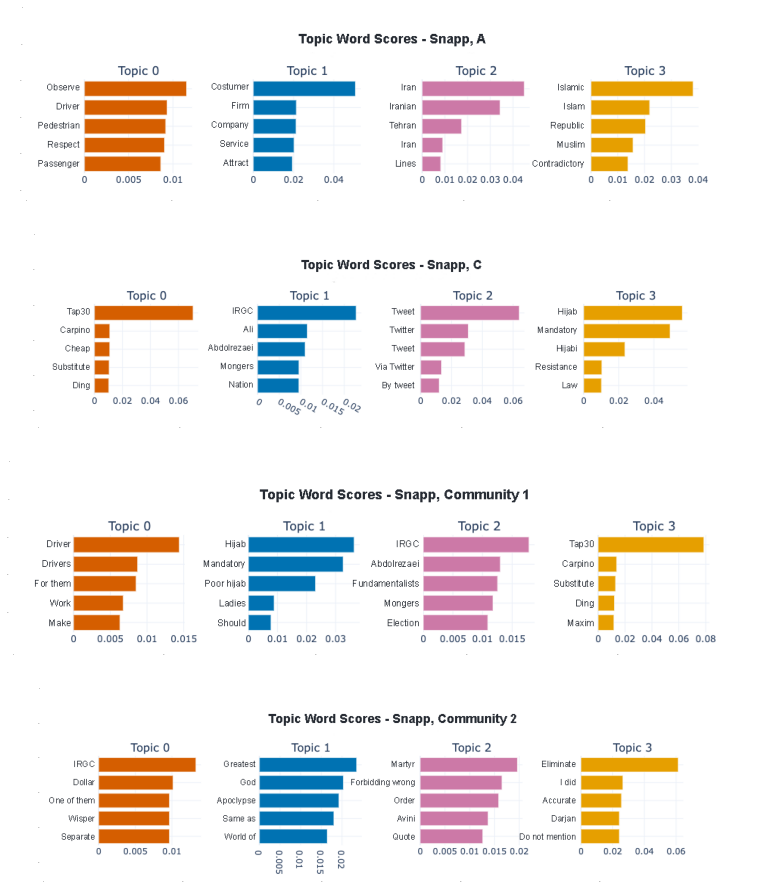}
	\caption{ \footnotesize { Salient topics and representative words, obtained from BERTopic, for tweets from low-CAP (C) and high-CAP (A) on the top two rows and from user in two major friendship communities on the bottom two rows, for the Snapp discussion. 
	}}
	\captionsetup{labelformat=empty}
	\label{appendix_fig:topics_snapp}
\end{figure}

\begin{figure}[hbt!]
	\centering
	\captionsetup{width=.9\linewidth, format=hang}
	\includegraphics[width=.85\linewidth]{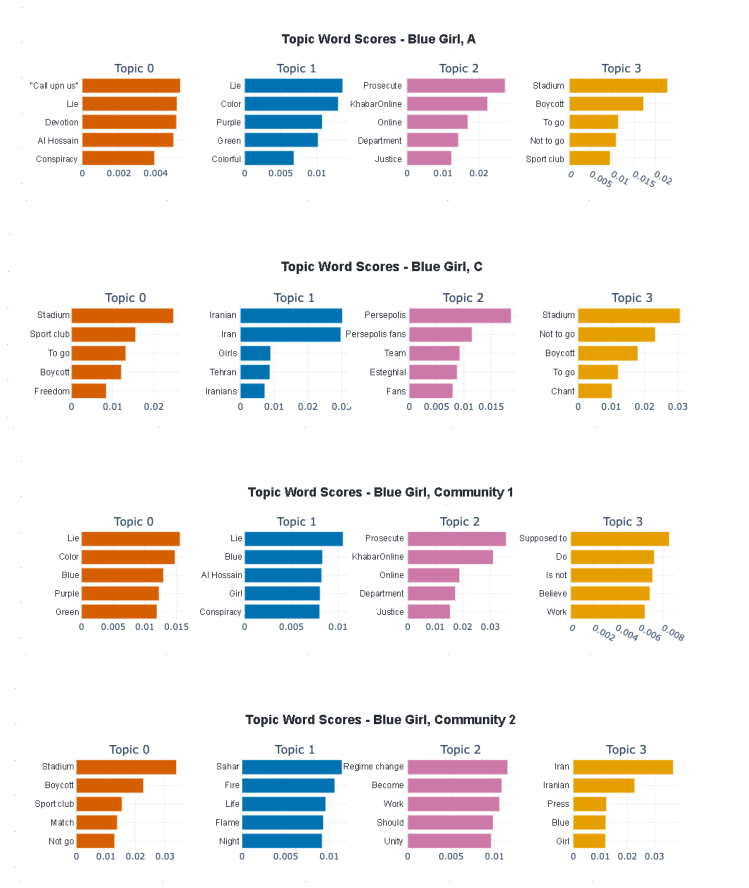}
	\caption{ \footnotesize { Salient topics and representative words, obtained from BERTopic, for tweets from low-CAP (C) and high-CAP (A) on the top two rows and from user in two major friendship communities on the bottom two rows, for the Blue Girl discussion. 
	}}
	\captionsetup{labelformat=empty}
	\label{appendix_fig:topics_bluegirl}
\end{figure}

\begin{figure}[hbt!]
	\centering
	\captionsetup{width=.9\linewidth, format=hang}
	\includegraphics[width=.85\linewidth]{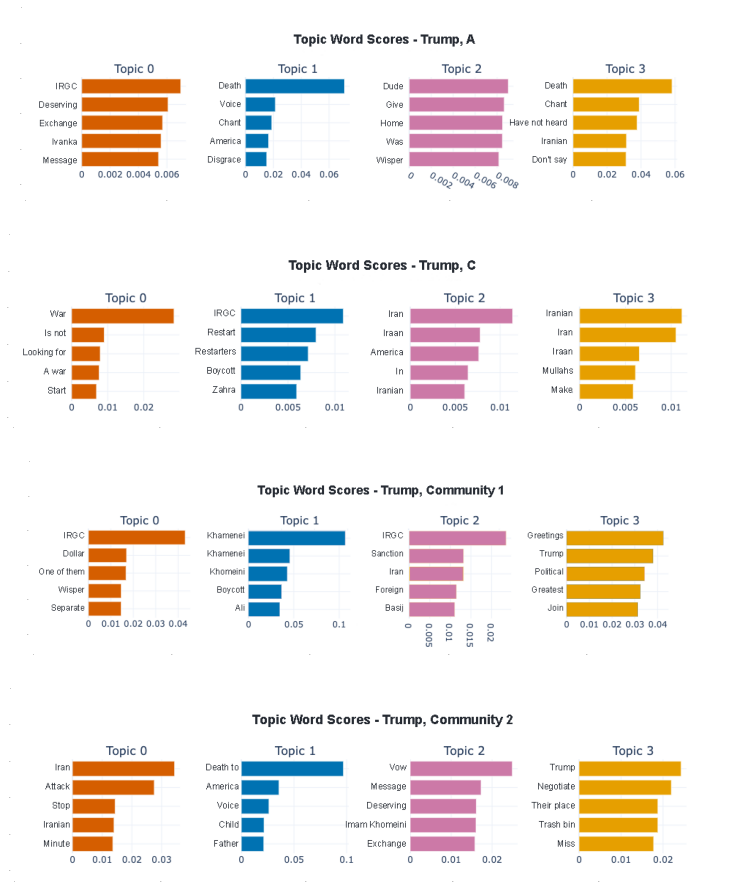}
	\caption{ \footnotesize { Salient topics and representative words, obtained from BERTopic, for tweets from low-CAP (C) and high-CAP (A) on the top two rows and from user in two major friendship communities on the bottom two rows, for the Trump discussion. 
	}}
	\captionsetup{labelformat=empty}
	\label{appendix_fig:topics_trump}
\end{figure}

\begin{figure}[hbt!]
	\centering
	\captionsetup{width=.9\linewidth, format=hang}
	\includegraphics[width=.85\linewidth]{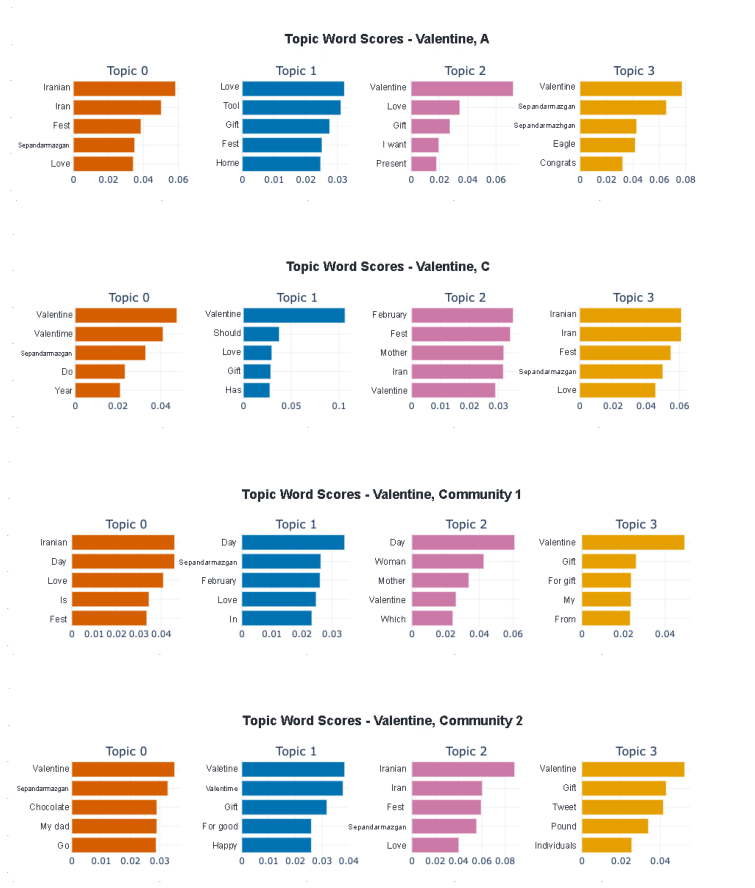}
	\caption{ \footnotesize { Salient topics and representative words, obtained from BERTopic, for tweets from low-CAP (C) and high-CAP (A) on the top two rows and from user in two major friendship communities on the bottom two rows, for the Valentine discussion. 
	}}
	\captionsetup{labelformat=empty}
	\label{appendix_fig:topics_valentine}
\end{figure}

\begin{figure}[hbt!]
	\centering
	\captionsetup{width=.9\linewidth, format=hang}
	\includegraphics[width=.85\linewidth]{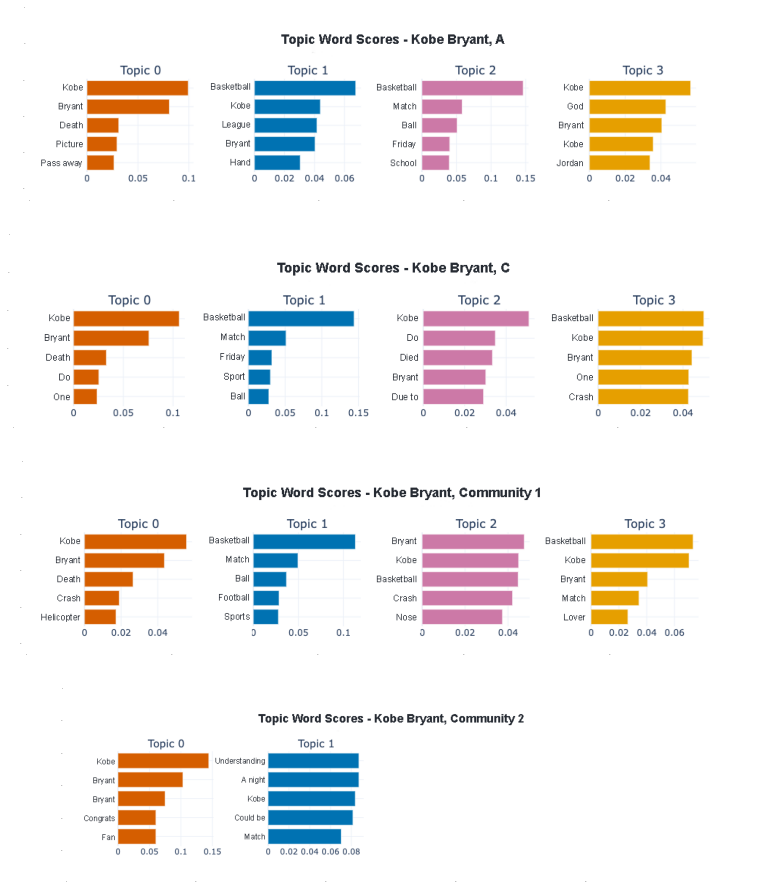}
	\caption{ \footnotesize { Salient topics and representative words, obtained from BERTopic, for tweets from low-CAP (C) and high-CAP (A) on the top two rows and from user in two major friendship communities on the bottom two rows, for the Kobe Bryant discussion. 
	}}
	\captionsetup{labelformat=empty}
	\label{appendix_fig:topics_kobe_bryant}
\end{figure}

\begin{figure}[hbt!]
	\centering
	\captionsetup{width=.9\linewidth, format=hang}
	\includegraphics[width=.85\linewidth]{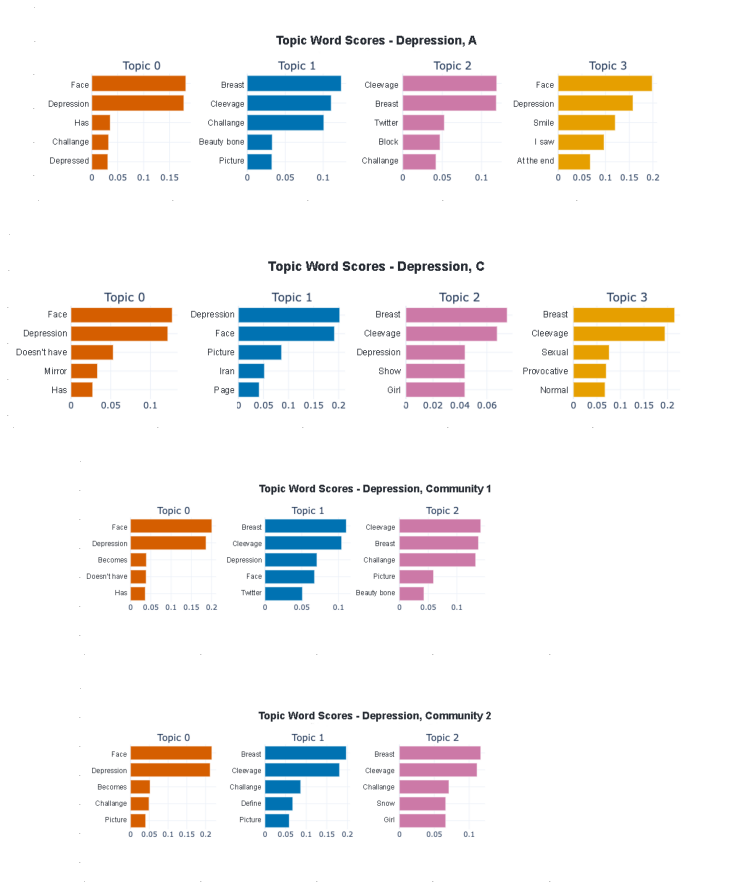}
	\caption{ \footnotesize { Salient topics and representative words, obtained from BERTopic, for tweets from low-CAP (C) and high-CAP (A) on the top two rows and from user in two major friendship communities on the bottom two rows, for the Depression discussion. 
	}}
	\captionsetup{labelformat=empty}
	\label{appendix_fig:topics_afsordegi}
\end{figure}

\begin{figure}[hbt!]
	\centering
	\captionsetup{width=.9\linewidth, format=hang}
	\includegraphics[width=.85\linewidth]{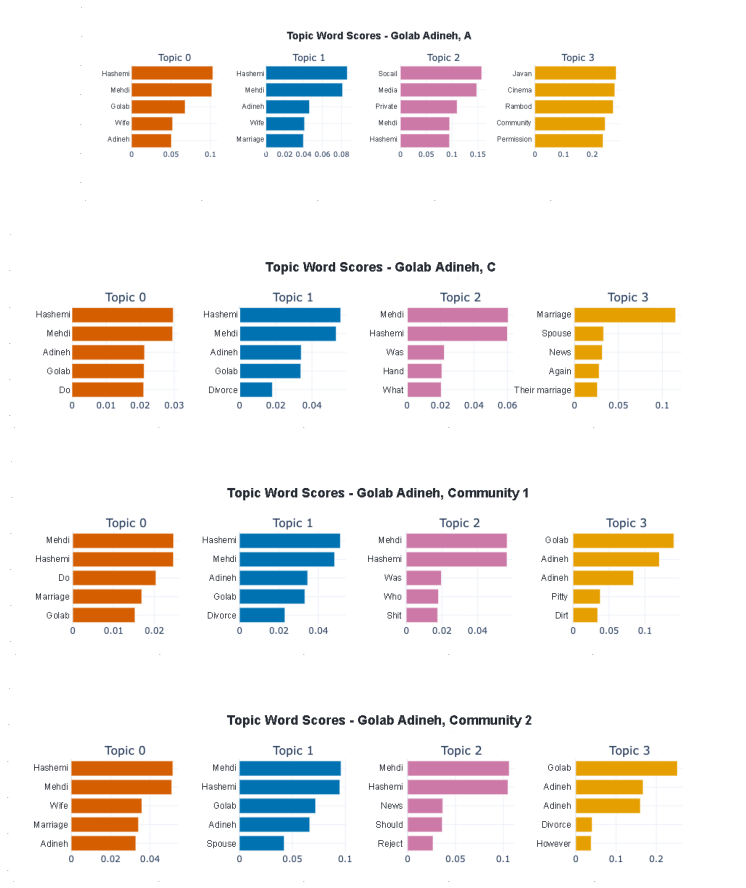}
	\caption{ \footnotesize { Salient topics and representative words, obtained from BERTopic, for tweets from low-CAP (C) and high-CAP (A) on the top two rows and from user in two major friendship communities on the bottom two rows, for the Golab Adineh discussion. 
	}}
	\captionsetup{labelformat=empty}
	\label{appendix_fig:topics_golab_adine}
\end{figure}

{
Consistent with other results reported, for the differences in the content of tweets posted by users from different user types and network communities, in the Results section of the main manuscript we report the results for the Snapp and Valentine discussions as representatives of divisive political and apolitical discussions.
In this part of this appendix, we include the word clouds as well as detected topics with their representative words for other divisive political and apolitical discussions, constructed as described in the Results and Methods sections of the main manuscript.
The observations in Figures \ref{appendix_fig:wordclouds_1vs2} through \ref{appendix_fig:topics_golab_adine} confirm that the content analysis discussed in the main manuscript can be generally extended to other discussions in the two target categories.
Note that the word clouds for three apolitical discussions were empty showing no difference between the two communities, and hence were not included in Fig. \ref{appendix_fig:wordclouds_1vs2}.
}

    \clearpage
    
    \section{Topics of discussion\zlabel{Appendix_topics}}
    
    In this appendix we explain the topics of discussion we chose for this paper.
We analyzed the discussion space of the following eleven topics:
\begin{itemize}
    \item \textbf{Bahareh Hedayat:} 
          This discussion started after the arrest of Bahareh Hedayat - a student activist at the University of Tehran.
          The discussion was limited in scale as the news did not find its way to mainstream media.
          Nevertheless, Twitter accounts from both sides of the political spectrum tweeted about it.
          The topic is political, but the dominant voice belonged to groups opposing the arrest - including some Iranian officials affiliated with the government or journalists active in Iran.
          Therefore, this is a non-divisive political discussion.
          Our data for this discussion contains 9794 tweets from 4651 unique user IDs.
          
    \item \textbf{Blue Girl:} 
          This discussion took place after a female soccer fan in Iran committed political self-immolation in public in protest to the existing ban on women entering soccer stadiums.
          The discussion was one of the main trending discussions at the time, and users from across the political spectrum tweeted about the issue.
          Since the more conservative groups in Iran believe that the ban is in the interest of the public, the discussion space was quite polarized.
          This discussion is a divisive political discussion.
          Our data for this discussion contains 539682 tweets from 57616 unique user IDs.
          
    \item \textbf{Depression:} 
          This topic is endemic to Twitter.
          A user posted her selfie with the caption \textit{"This is the face of depression"}.
          Since the user---a woman--- was wearing revealing clothes in the picture, it was deemed provocative and the intention of the user was questioned by Farsi Twitter users.
          This was followed by both a series of parody selfies and a series of serious tweets on the issue of depression.
          This discussion remained mainly social and apolitical. 
          Our data for this discussion contains 5016 tweets from 3943 unique user IDs.

    \item \textbf{Drone:} 
          This discussion followed the downing of the American drone in the Persian Gulf in 2019.
          The event was a part of a sequence of events heightening the tensions between the governments of Iran and the United States.
          Many Iranians were genuinely concerned about a seemingly imminent conflict.
          The discussion space was dominated by the supporters of the Iranian government praising the Iranian military for its response.
          Many dissident groups did not participate in this discussion and, although political, the discussion was mostly non-divisive.
          Our data for this discussion contains 40201 tweets from 15916 unique user IDs.

    \item \textbf{Golab Adineh:} 
          This discussion started after the news about a controversial celebrity marriage, following a celebrity divorce, broke out.
          Mehdi Hashemi, a renowned Iranian actor, then 72 years old, married Mehnoosh Sadeghi, an actress who was at the time 46 years old.
          Mehdi Hashemi was previously married to Golab Adineh, a then 65 year old famous actress, who was separated from Mehdi Hashemi for a while before their recent divorce.
          The discussion sparked a social controversy about second marriage and women's rights.
          Although it was a heated-up social discussion, it never became political.
          Our data for this discussion contains 5962 tweets from 4117 unique user IDs.

    \item \textbf{Kobe Bryant:} 
          A series of tweets from a variety of accounts followed the death of the NBA athlete, Kobe Bryant.
          Although similar attentions to such topics linked to western culture is typically potent of sparking controversy, with possible ideological or political flavor, this discussion remained apolitical.
          Our data for this discussion contains 3662 tweets from 2872 unique user IDs.

    \item \textbf{Plane:}
          Following the downing of the Ukrainian plane in Iran, which itself followed the assassination of Qasem Soleimani, Farsi Twitter became a vibrant scene of political discussion about the incident.
          Since no group or individual approved of the downing of the plane, users on one end of the political spectrum remained mostly silent and the discussion space was dominated by those opposing the Iranian government and condemning the action. 
          Although in this respect this discussion shares characteristics with non-divisive political discussions, given the unique context with highly divisive prelude and consequences, this discussion stands out as an isolated case per se.
          Our data for this discussion contains 123614 tweets from 37552 unique user IDs.

    \item \textbf{Snapp:}
          A driver of the most popular ride-sharing app in Iran, Snapp, asked his passenger to cover her hair with her headscarf in observation of the existing Hijab laws in Iran making wearing a headscarf mandatory for women.
          Following the passenger's refusal to comply, the driver terminated the ride and asked her to leave the car.
          The passenger posted her experience on social media and publicly asked the ride-sharing company, Snapp, to be responsible for their driver's behavior.
          The company initially sided with the customer, but following a series of tweets by the more conservative groups supporting the driver, they retracted their initial statement and supported the driver.
          The state TV and a number government officials spoke out about the incident supporting the driver's stance and praised him for trying to uphold the hijab law.
          This was followed by a reaction from the groups that oppose mandatory hijab, some of whom directly targeted the Iranian government and pivoted into discussing more general issues about social freedom, women's rights, and the general approach of the Iranian government towards gender equality.  
          The discussion became extremely polarized and remained a divisive political discussion.
          Our data for this discussion contains 144347 tweets from 33253 unique user IDs.

    \item \textbf{Suicide:}
          Saeed Namaki, the then Iranian Minster of Health and a relatively progressive figure in the Iranian political scene, made a comment about suicide and depression in Iran.
          This coincided with a recent suicide that made its way to the news and triggered a discussion on Twitter.
          Since a political figure was involved, some users linked depression in Iran to political issues and the discussion became political, while being simultaneously a social discussion.
          It was mainly non-divisive.
          Our data for this discussion contains 112783 tweets from 41665 unique user IDs.

    \item \textbf{Trump:}
          In the height of the tension between the governments of the United States and Iran in the spring and summer of 2019, following the attack on the oil tankers in the Persian Gulf, the Aramco attack, the downing of the American drone, and tightened sanctions, the US president \textit{Donald Trump} became a trending topic of discussion in Iran.
          Although the related issues were less divisive, a considerable population of users in Farsi twitter who were against the Iranian government chose to support the stance of Trump's administration.
          On the other hand, the supporters of the Iranian government and a noticeable part of those opposing the government, were against Trump.
          Therefore, this discussion became a politically divisive one. 
          Note that this discussion, unlike the others, had more than one peak.
          The data was collected over a long period and there were multiple peaks in the trendiness of the discussion.
          Our data for this discussion contains 112864 tweets from 29416 unique user IDs.

    \item \textbf{Valentine:}
          Before the Valentine's day in 2019, discussions about Valentine became trending in Farsi Twitter and remained trending until a few days after the Valentine's day.
          Since Valentine is not traditionally observed in Iran, it is considered an imported western tradition and the conservative groups of the society stand against its celebration.
          Meanwhile, it is widely observed by many groups of the urban population in Iran who support the celebration of this day.
          Despite its potentials for becoming political, this discussion remained apolitical.
          Our data for this discussion contains 7962 tweets from 5424 unique user IDs.

\end{itemize}

\clearpage

\end{appendices}

\end{document}